\documentclass[iop]{emulateapj}
\usepackage{amsmath}
\usepackage{units}
\usepackage[titletoc,title]{appendix}
\usepackage{mathtools}

\newcommand{\fid}[3] {\left( \frac{#1}{#2} \right)^{#3}}

\graphicspath{{../figures/}{../../Thesis_Code/gas_assist_plots}}
\usepackage[normalem]{ulem}
\usepackage[breaklinks,colorlinks,urlcolor=cyan,citecolor=cyan,linkcolor=cyan]{hyperref}
\usepackage{graphicx}

\begin{document}
	
	\title{Restrictions on the Growth of Gas Giant Cores via Pebble Accretion}
	\author{M. M. Rosenthal}
	\author{R. A. Murray-Clay}
	\affil{Department of Astronomy and Astrophysics, University of California, Santa Cruz, CA 95064, USA}
	\date{\today}
	
\begin{abstract}
We apply an order-of-magnitude model of gas-assisted growth, known as pebble accretion, in a turbulent medium to suggest a reason why some systems form wide orbital separation gas giants while others do not. In contrast to traditional growth by ballistic collisions with planetesimals, growth by pebble accretion is not necessarily limited by doubling times at the highest core mass. Turbulence, in particular, can cause growth to bottleneck at lower core masses. We demonstrate how a combination of growth by planetesimal and pebble accretion limits the maximum semi-major axis where gas giants can form. We find that, for fiducial disk parameters, strong turbulence ($\alpha \gtrsim 10^{-2}$) restricts gas giant cores to form interior to $a \lesssim 40 \, \text{AU}$, while for weak turbulence gas giants can form out to $a \lesssim 70 \, \text{AU}$. The correspondence between $\alpha$ and semi-major axis depends on the sizes of small bodies available for growth. This dependence on turbulence and small-body size distribution may explain the paucity of wide orbital separation gas giants. We also show that while lower levels of turbulence ($\alpha \lesssim 10^{-4}$) can produce gas giants far out in the disk, we expect these gas giants to be low-mass ($M \lesssim \, 1 M_J$). These planets are not luminous enough to have been observed with the current generation of direct-imaging surveys, which could explain why wide orbital separation gas giants are currently observed only around A stars.
\end{abstract}

\section{Introduction}
In the traditional ``core accretion" model of planet formation, growth of planets proceeds in a bottom-up manner. Planets begin their growth as rocky cores, or protoplanets. If these protoplanets reach sufficient size within the lifetime of the gas disk, they will be able to trigger runaway gas accretion, resulting in a gas giant (\citealt{pollack_gas_giants}). This runaway occurs when $M_{\rm{atm}} \sim M_{\rm{core}}$, where $M_{\rm{atm}}$ is the mass of the planet's atmosphere and $M_{\rm{core}}$ is the mass of the planet in solids. The critical core mass, $M_{\rm{crit}}$, where this occurs is usually quoted as $M_{\rm{crit}} \sim 10 M_\oplus$, though the actual mass depends on the disk parameters, especially the opacity and the core's accretion rate (see, e.g. \citealt{raf06}, \citealt{pymc_2015}). A gas giant will not form if the planet cannot reach $M_{\rm{crit}}$ within the lifetime of the gas disk, $\tau_{\rm{disk}}$, which is $\sim 2.5 \, \text{Myr}$ for G stars (\citealt{lifetimes_mamajek}, \citealt{lifetimes_ribas}). Traditional models rely on gravitational focusing to increase the effective radius for collisions. These models, which we will refer to as ``canonical core accretion" or ``planetesimal accretion" models, give growth timescales that are generally fast enough to reach critical core mass for $a\lesssim10\,\text{AU}$, but become longer than the disk dispersal timescale past this distance. (See \citealt{gold}, hereafter GLS, for a review of gas-free regimes.) 

Observations of exoplanetary systems have challenged this canonical core accretion model in a number of ways. Here we focus on the existence of systems that feature gas giants at wide orbital separations (see, e.g. \citealt{bowler_DI_review} for a review). Of particular note is the planetary system surrounding the star HR 8799, which exhibits a nonhierarchical, multiplanet structure:  HR 8799 consists of four gas giant planets ($M\sim10 \, M_J$)  at extremely wide projected separations: 14, 24, 38, and 68 AU (\citealt{HR8799_orig}, \citealt{HR8799_fourth}). HR 8799 poses a serious challenge to canonical core accretion models because the last doubling timescale for growth at these distances is far too long for a core to reach the critical mass necessary to trigger runaway growth within $\tau_{\rm{disk}}$. Additional effects, such as gas drag from the planet's atmosphere (\citealt{ii_03}) or damping of the planetesimals' random motions by the nebular gas (\citealt{raf}), can increase the cross section for collisions further. Neither of these effects, however, are sufficient to allow the \textit{in situ} formation of gas giants at $70 \, \text{AU}$.

A number of alternative formation scenarios have been proposed to explain the formation of HR 8799. One commonly suggested explanation is that HR 8799 is evidence of an alternative formation scenario known as ``gravitational instability," wherein the gaseous component of the protoplanetary disk becomes unstable to gravitational collapse and subsequently fragments into the observed gas giant planets (\citealt{boss_1997}; see also \citealt{kl_2016} for a more recent review). However, \cite{kratter_gas_giants} pointed out that it is difficult to form fragments of the sizes seen in HR 8799 without having these ``planets" grow to brown dwarf or even M-star masses. The lack of observed brown dwarfs at wide orbital separations provides some evidence against this hypothesis, but additional statistical work is needed (\citealt{bowler_DI_review}). Outward scattering after formation at smaller orbital separations is another possibility, but $N$-body simulations by \cite{dodson-robinson_gas_giants} find that scattering is unlikely to produce systems with the multiplanet architecture of HR 8799.

In recent years, a third possibility has emerged: a modification to the theory of core accretion commonly referred to as ``pebble accretion," which we will also refer to as ``gas-assisted growth"  (\citealt{OK10}, \citealt{pmc11}, \citealt{OK12}, \citealt{lj12}, \citealt{LJ14}, \citealt{lkd_2015}, \citealt{mljb15}, \citealt{vo_2016}, \citealt{igm16}, \citealt{xbmc_2017}, \citealt{rmp_2018}). In pebble accretion, the interaction between solid bodies and the gas disk is considered in detail when determining the growth rates of planets. In particular, gas drag can enhance growth rates by removing energy from small bodies. Particles that deplete their kinetic energy within the gravitational sphere of influence of a larger body can become bound to this parent body, which will eventually lead to accretion of the particle by the growing protoplanet. This process can occur at larger impact parameters than are required for the particle to collide with the core, which in turn increases the accretion cross section. This interaction often affects mm-cm-sized bodies the most strongly. Note, however, that for low-density, ``fluffy" aggregates, the radius of bodies most substantially affected by gas drag  can be substantially larger. 

\begin{figure} [h]
	\centering
	\includegraphics[width=\linewidth]{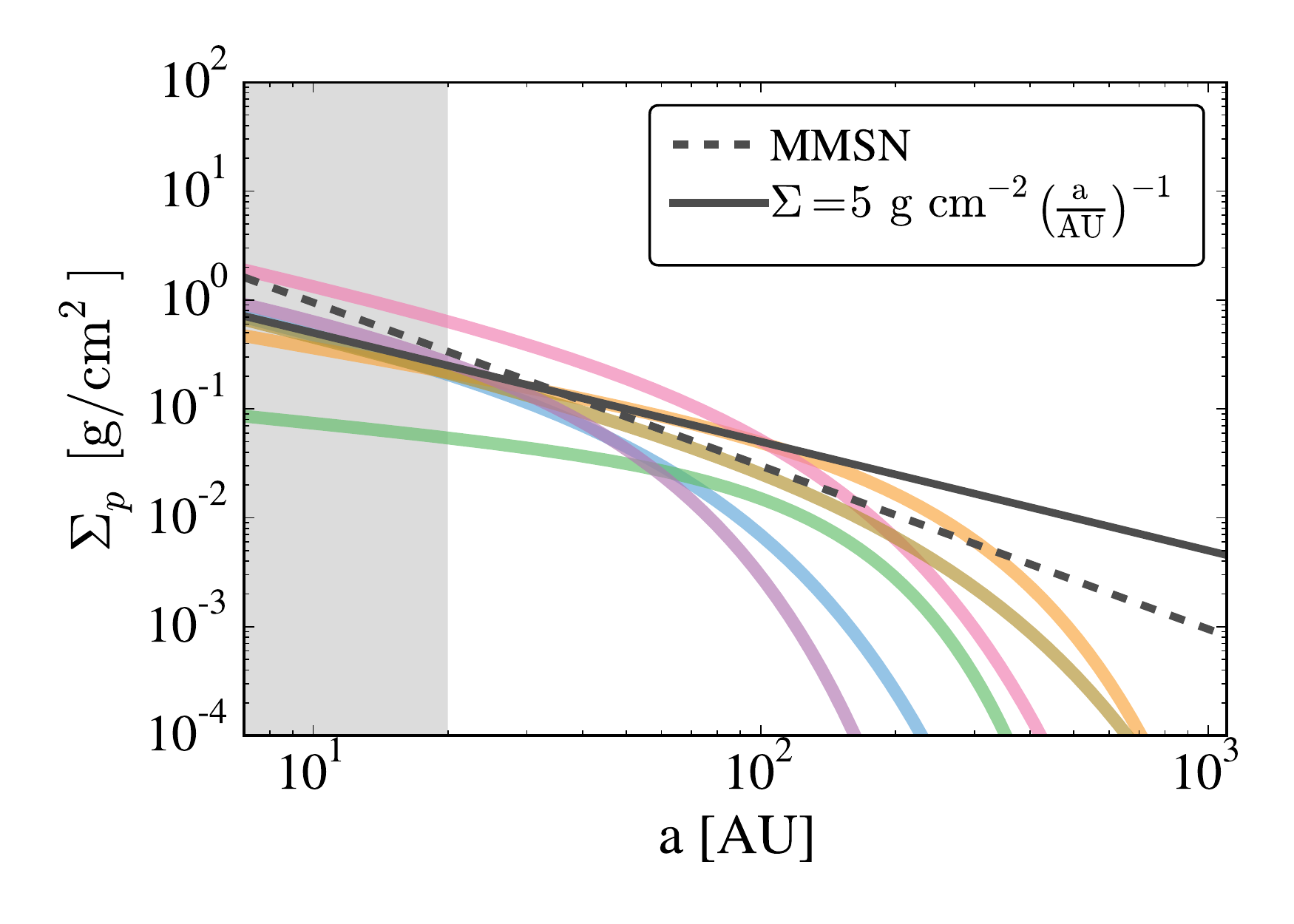}
	\caption{Colored lines show the dust surface density in $0.1 \, \text{mm} - 1 \, \text{mm}$ sized particles taken from $870$ $\mu \text{m}$ continuum emission observations of protoplanetary disks done by \cite{andrews_09}. See text for details. Also shown for reference is the value of the solid surface density in the minimum-mass solar nebula (MMSN), appropriate for the outer disk, $30 \, (a/\text{AU})^{-3/2} \, \text{g} \, \text{cm}^{-2}$ (\citealt{weid_mmsn}, \citealt{hay_mmsn}), as well as the fiducial surface density used in this work to match the observations. In the gray shaded region the values of the curves are extrapolations to scales smaller than the observations can resolve.}
	\label{fig:andrews09}
\end{figure}

For gas-assisted growth to operate, a reservoir of pebble-sized objects must exist in the protoplanetary disk. Because the sizes of these pebbles are comparable to the $\sim$ mm wavelengths used to measure dust surface densities in the outer regions of protoplanetary disks, observations can directly probe the surface densities in the small solids that fuel gas-assisted growth. These observations find large reservoirs of small, pebble-sized solids (\citealt{andrews_09}, \citealt{andrews}). An example is shown in Figure \ref{fig:andrews09}, which presents disk surface densities measured by \cite{andrews_09}. The figure shows the surface density in particles of radius $0.1 \, \text{mm} - 1 \, \text{mm}$, which is inferred by integrating the size distribution used in the paper ($d N / d r_s \propto r_s^{-3.5}$) from 0.1 to $1 \, \text{mm}$. Performing the integration gives the fraction of the measured solid surface density contained in this size range ($\sim 70\%$).

Given this observed reservoir of small solids, pebble accretion dramatically increases the expected growth rate of large cores. Under fiducial conditions, the timescale for a core's last doubling to canonical values of $M_{\rm{crit}}$ is below the disk lifetime, even at many tens of AU separations. Though fast accretion of solids deposits enough energy to delay the onset of runaway accretion of a gas envelope, once a core has reached several Earth masses, finely tuned disk conditions are required to slow atmospheric growth enough to prevent runaway from ultimately occurring. Thus, growth via pebble accretion seems to predict that wide orbital separation gas giants should be common. However, direct imaging surveys show that planets $\gtrsim 2-5 \, M_J$ are rare at distances $> 30 \, \text{AU}$ (\citealt{brandt_di}, \citealt{chauvin_di}, \citealt{bowler_DI_review}, \citealt{gal_gas_giant_freq}). 

One possibility for solving this problem is the presence of turbulence in the nebular gas. In this work, by ``turbulence," we generally mean any anomalous root mean square (RMS) velocity of the nebular gas that is not due to the laminar velocity that arises from radial pressure support in the disk. The main effect of turbulence on pebble accretion is to increase the velocity dispersion of the pebbles due to their coupling with the gas; it is only in Section \ref{final_mass} that we connect our parameterization of the turbulent RMS velocity to the transport of angular momentum in the disk. Turbulence can both increase the kinetic energy of an incoming particle and decrease the core's gravitational sphere of influence. Turbulence also drives particles vertically, reducing the overall densities of small bodies and slowing accretion. Turbulence is usually only included in models of protoplanetary growth by pebble accretion by increasing the particle scale height and hence reducing the mass density of solids. Some models of the early stages of planetesimal growth discuss the effects of turbulence (e.g. \citealt{gio_2014}, \citealt{hgb_2016}), but these models are concerned with accretion at cross sections comparable to the core's geometric cross section; i.e. they neglect the effects of the core's gravity.

In this paper, we use an order-of-magnitude model of pebble accretion (\citealt{rmp_2018}, hereafter R18) to propose a criterion for the formation of gas giants via gas-assisted growth. In particular, R18 investigated how turbulence affects the growth of gas giant cores as a function of core mass.  High-mass cores ($\gtrsim 10^{-2}-10^{-1} M_\oplus$) can grow on timescales less than the lifetime of the gas disk, even in strong turbulence. However, for lower-mass cores and stronger turbulence, the range of pebble sizes available for growth is restricted. In this case, the pebble sizes for which growth is most efficient often cannot be accreted, and growth can ``stall" at low core masses.

In effect, a core must first achieve a minimum mass before it can quickly grow to $M_{\rm{crit}}$ via gas-assisted growth. In this paper, for our fiducial calculation, we assume that growth to this minimum mass happens by canonical core accretion, which allows us to place semi-major axis limits on where gas giant growth is possible. We also calculate values for the core mass needed at a given semi-major axis for pebble accretion to be rapid, which apply regardless of how the early stages of growth proceed. The assumption that low-mass growth is fueled by planetesimal accretion requires that, in addition to the reservoir of small pebbles, a substantial population of larger planetesimals has formed. We discuss the ramifications of varying the mass in planetesimals in Section \ref{upper_lim}. Close to the central star, planetesimal accretion can dominate the early growth of planets, with pebble accretion setting the growth timescale for high-mass cores. Far from the central star, however, planetesimal accretion is less efficient, limiting its ability to grow cores to high enough masses that pebble accretion kicks in. Thus, turbulence can set the maximum distance at which gas giant formation is possible via pebble accretion. We find that for quiescent disks, gas giants can form far out in the disk ($a \lesssim 70 \, \text{AU}$), but for stronger turbulence, this maximum distance is smaller (e.g. $a \lesssim 40 \, \text{AU}$ for $\alpha \gtrsim 10^{-2}$). Furthermore, while disks with weaker turbulence can have gas giants at wider orbital separation, the weaker viscosities in these disks mean that the masses of the gas giants formed are likely lower ($\lesssim 2 \, M_J$), which would preclude them from being detected by the current generation of direct-imaging surveys. Therefore, there may exist a population of wide orbital separation gas giants that have yet to be found due to their low luminosities.

In Section \ref{overview}, we review our model, which is discussed in detail in R18. In Section \ref{wide_sep}, we discuss how gas-assisted growth operates at wide orbital separation, contrasting the rapid growth at high core mass with the slower growth for low-mass cores. In Section \ref{gas_giants}, we explore how turbulence can place limits on the semi-major axes where gas giants can form. In Section \ref{final_mass}, we investigate the implications for the final masses of gas giants if turbulence plays a role in gap opening in addition to early core growth. Finally, in Section \ref{summary}, we summarize our results and give our conclusions.

\section{Model Overview} \label{overview}

In this section, we will give a brief summary of the ideas behind pebble accretion and how they are implemented in our model. We will focus on pebble accretion at the mass scales relevant to limiting gas giant growth -- i.e. masses in the range $10^{-4} M_\oplus \lesssim M \lesssim 10^{-2} M_\oplus$ (see Figure \ref{fig:m_min}). A more general and in-depth discussion can be found in the Appendix, and in R18.

\subsection{Basic Pebble Accretion Processes} \label{basic}

In this section, we discuss the basic parameters that go into calculating the growth timescale and contrast gas-assisted growth with growth via planetesimal accretion.

The setup for our model consists of a large body, or protoplanetary ``core," growing by accreting a population of small bodies. Our calculation is performed for a given size of small body, expressed either in terms of the small body's mass, $m$, or its radius $r_s$. Note that, practically speaking, the important parameter for our calculation is the particle's Stokes number, $St$ (see Section \ref{t_grow}). We can convert from Stokes number to radius or mass by assuming a density for the small bodies. In what follows, we will assume a density of $\rho_s = 2 \, \text{g} \, \text{cm}^{-3}$, which is appropriate for rocky or icy bodies. We note, however, that lower density, fluffy aggregates will have higher radii at a given Stokes number.

In general, the growth timescale for the large body of mass $M$ is given by

\begin{align}
t_{\rm{grow}} = \left(\frac{1}{M}\frac{dM}{dt}\right)^{-1},
\end{align}
while the growth rate, $dM/dt$ can be expressed as

\begin{align} \label{eq:m_dot}
\frac{dM}{dt} = m (n \sigma_{\rm{acc}} v_\infty) = m \left(\frac{f_s \Sigma}{2 H_p m} \right) (2 R_{\rm{acc}}) (2 H_{\rm{acc}}) v_\infty \, .
\end{align}
Here $n$ is the volumetric number density of small bodies, $\sigma_{\rm{acc}}$ is the accretion cross section, and $v_\infty$ is the velocity at which small bodies approach the large body. In the second equality, we have set $n=f_s \Sigma / (2 H_p m)$, where $H_p$ is the scale height of the small bodies, $\Sigma$ is the surface density of the gas, and $f_s\equiv\Sigma_p/\Sigma$ is the solid-to-gas mass ratio in the disk. We have also decomposed $\sigma_{\rm{acc}}$ into the product of length scales parallel and perpendicular to the disk plane, $2 R_{\rm{acc}}$ and $2 H_{\rm{acc}}$, respectively. Combining these two expressions gives

\begin{align} \label{eq:t_grow}
t_{\rm{grow}} = \frac{M H_p}{2 f_s \Sigma v_\infty R_{\rm{acc}} H_{\rm{acc}}} \; .
\end{align}
Thus, once $H_p$, $v_\infty$, $R_{\rm{acc}}$, and $H_{\rm{acc}}$ are determined, $t_{\rm{grow}}$ can be calculated.

For growth that proceeds by accretion of massive planetesimals, the effects of gas drag are generally negligible (though see \citealt{raf} for a discussion of the effects of gas drag on smaller planetesimals of size $\lesssim 1 \, \rm{km}$). In this case, the value of $R_{\rm{acc}}$ is determined by the maximum impact parameter at which a small body will be gravitationally focused into a collision with the core,

\begin{align} 
R_{\rm{focus}}  = R \left( 1  + \frac{v_{\rm{esc}}^2}{v_{\infty}^2} \right )^{1/2} \; ,
\end{align}
where $R$ is the physical radius of the core, and $v_{\rm{esc}} = \sqrt{2 G M /R}$ is the escape velocity from the core.

An important parameter for calculating $R_{\rm{focus}}$ is the core's ``Hill radius," which is the characteristic radius at which the large body's gravity strongly influences the trajectories of the small bodies.   For a big body of mass $M$ orbiting a star of mass $M_*$ at a semi major axis $a$, the Hill radius, $R_H$ is given by (\citealt{hill}),

\begin{align}
R_H = a \left( \frac{M}{3 M_*} \right)^{1/3} \; ,
\end{align}
which can be obtained by determining the length at which the gravity of the large body is equal to the tidal gravity from the central star. Particles that pass within distances $\sim$ $R_H$ of the core move on complex trajectories that cannot be expressed as a simple function of impact parameter (\citealt{hill_enc}). Particles that emerge from the Hill radius without colliding with the large body will generally have their velocities relative to the core excited up to $v_\infty \sim R_H \Omega \equiv v_H$ in a random direction (GLS), where $\Omega = \sqrt{G M_* / a^3}$ is the Keplerian angular frequency and $a$ is the semi-major axis of the core's orbit. The quantity $v_H$ is known as the ``Hill velocity." If $v_\infty \sim v_H \ll v_{\rm{esc}}$, it is straightforward to show that
\begin{align} \label{eq:r_focus}
R_{\rm{focus}} \sim \sqrt{R R_H} \; .\end{align} 
Since interactions with the core excite planetesimals to a random velocity $v_\infty \sim v_H$, this is the largest capture radius possible for planetesimal accretion without invoking some damping mechanism to lower the planetesimal velocity below $v_H$. Note, however that since $R \ll R_H$, $R_{\rm{focus}} < R_H$.

In gas-assisted growth, on the other hand, the value of $R_{\rm{acc}}$ can be much larger than $R_{\rm{focus}}$. For ``pebble-sized" small bodies, the interaction between the small bodies and the gas is important when calculating the accretion rate. In particular, gas drag can remove kinetic energy from the small body as it interacts with the growing core. If the work done by gas drag is sufficiently large, small bodies that otherwise would have merely been deflected by the core's gravity can become gravitationally bound to the core, further reducing their energy and causing them to inspiral and eventually be accreted by the core. This can dramatically increase the impact parameters at which accretion will occur. As discussed by, e.g. R18, in certain regions of parameter space, the core can accrete over the entirety of its Hill sphere, i.e. accretion proceeds with $R_{\rm{acc}} = R_H \gg R_{\rm{focus}}$. 

\subsection{Pebble Accretion at Different Particle Radii} \label{part_size}

The Hill radius represents the largest distance at which particles can be captured. However, not all sizes of particles can be captured at $R_H$. To fully characterize the scale at which pebbles are captured, we need to introduce two additional radii. 

\begin{figure} [h]
	\centering
	\includegraphics[width=1.0\linewidth]{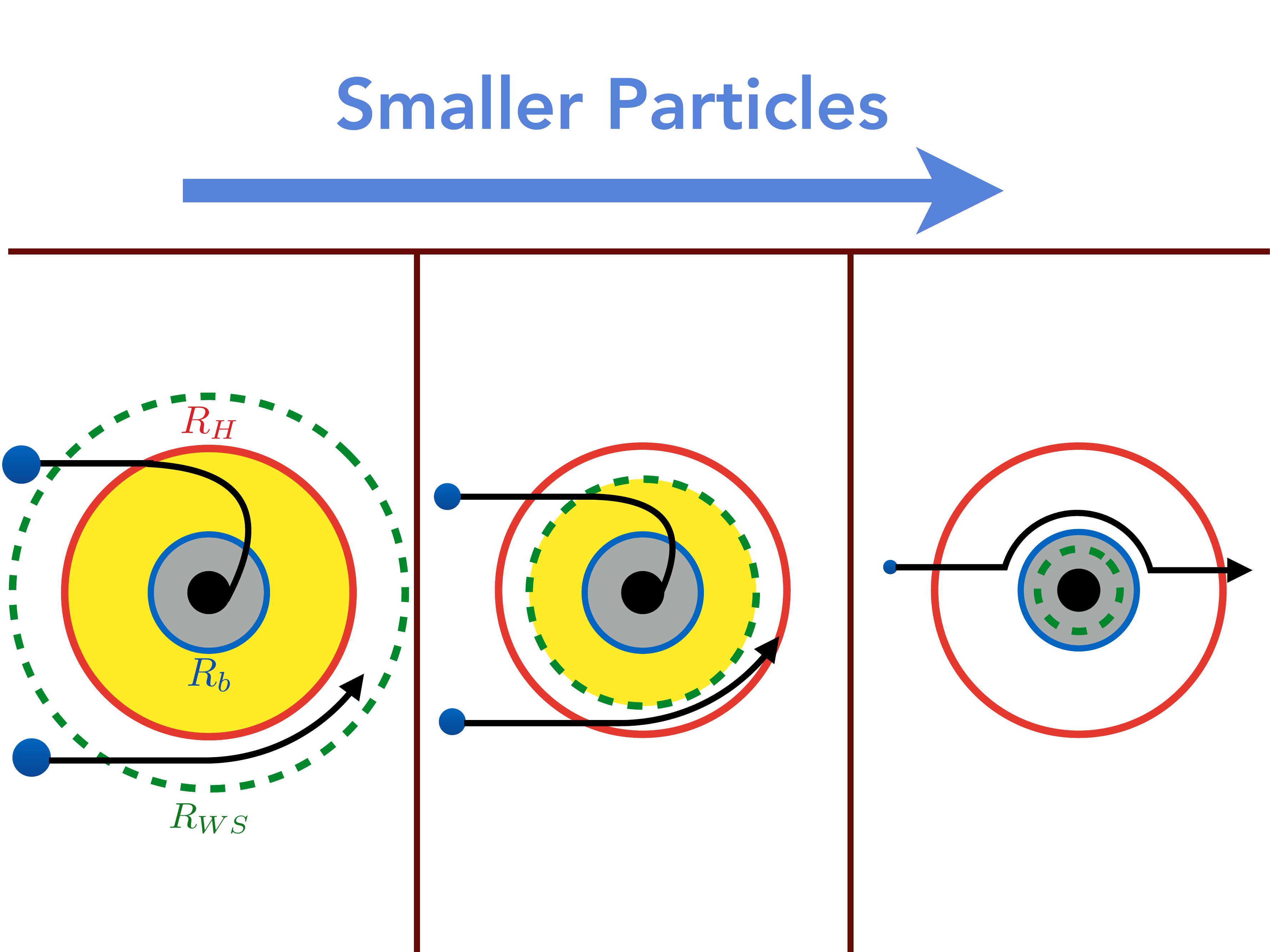}
	\caption{Illustration of how particle capture proceeds at different small-body radii. The central black circle represents the planet, and the blue circles represent incoming particles. The gray shaded region denotes the extent of the planet's atmosphere (or the planet's radius for $R>R_b$), and the yellow shaded region shows the region where incoming particles can be accreted. \textit{Left Panel}: for large particle radii, $R_{WS}^\prime > R_H$, so particles that enter $R_H$ and deplete their kinetic energy will be accreted by the core. Particles with impact parameters $<R_{WS}^\prime$ but $>R_H$ will not be able to accrete.  \textit{Middle Panel}: for intermediate sizes of particles, $R_{WS}^\prime < R_H$. These particles need to pass within $R_{WS}^\prime$ to be accreted, as if they pass within $R_H$ but at distances $>R_{WS}^\prime$ they will be sheared off the core by gas drag. \textit{Right Panel}: small particles will have $R_{WS}^\prime < R_b$. At this point, the core's gravitational sphere of influence is inside its atmosphere. Particles of these sizes will couple to the local gas flow, which will flow around the core's static atmosphere. Thus, particles of a size such that $R_{WS}^\prime < R_b$ will not be accreted via pebble accretion.}
	\label{fig:rs_ex}
\end{figure}

The first radius is the wind shearing (WISH) radius, which is the radius interior to which the core's gravity dominates over the differential acceleration between the small body and the core due to gas drag,
\begin{align}
R_{WS}^\prime = \sqrt{\frac{G(M+m)}{\Delta a_{WS}}} \; ,
\end{align}
where $\Delta a_{WS}$ is the differential acceleration between the two bodies due to gas drag (\citealt{pmc11}). Particles that approach the core at impact parameters $>R_{WS}^\prime$ will be pulled off the core by gas drag even if they are inside of $R_H$. Thus, the value of $R_{\rm{acc}}$ is given by
\begin{align}
R_{\rm{acc}} = \min(R_H,R_{WS}^\prime) \; .
\end{align}
However, the value of $R_{WS}^\prime$ depends on the size of the small body being accreted, unlike $R_H$. To see this, we note that if $M\gg m$, we can rewrite $R_{WS}^\prime$ as
\begin{align} \label{eq:r_ws_gen}
R_{WS}^\prime \approx \sqrt{\frac{G M t_s}{v_{\rm{rel}}}} \; .
\end{align}
Here $t_s$ is the stopping time of the small body,
\begin{align} \label{eq:t_s}
t_s \equiv \frac{m v_{\rm{rel}}}{F_D\left(m\right)} \; ,
\end{align}
$v_{\rm{rel}}$ is the relative velocity between the small body and the gas, and $F_D\left(m\right)$ is the drag force on the small body (see the Appendix for a discussion of how the correct $v_{\rm{rel}}$ for calculating $R_{WS}^\prime$ is determined). The stopping time parameterizes the size of the particle in terms of its interaction with the gas. Qualitatively, for large core masses only the stopping time of the smaller body is relevant because the core is essentially unaffected by gas drag. The largest particles that can deplete their kinetic energy will have $R_{WS}^\prime > R_H$, and will be able to accrete over the entirety of the core's Hill sphere, while smaller particles will have $R_{WS}^\prime < R_H$ and will only be accreted at more modest values of impact parameter. See the first two panels of Figure \ref{fig:rs_ex}. 

Pebble accretion will not continue down to arbitrary sizes of small bodies. As $R_{WS}^\prime$ decreases with decreasing particle size, we will eventually reach the scale of the core's atmosphere. Because the atmosphere is essentially static, and the flow velocity is subsonic, the local gas flow will not be able to penetrate into the core's atmosphere. Gas will instead flow around the static atmosphere held by the core. See, e.g. \cite{ormel_flows} for an example of this behavior in the context of a planet embedded in a protoplanetary disk. 

We take the scale of the core's atmosphere to be determined by the Bondi radius, $R_b$, which is the scale at which the escape velocity from the core is equal to the local isothermal sound speed $c_s = \sqrt{k T / \mu}$, where $k$ is Boltzmann's constant, $T$ is the temperature, and $\mu$ is the mean molecular weight of the gas molecules. Thus, $R_b$ is given by
\begin{align}
R_b = \frac{G M}{c_s^2} \; .
\end{align}
Once particles are small enough that $R_{WS}^\prime < R_b$, they need to penetrate into the core's atmosphere to become bound to the core. However, these small particles will couple strongly to the gas, which will flow around $R_b$, stopping the particles from accreting. Thus, we take $R_{WS}^\prime = R_b$ to set the smallest size of particles that can be accreted; see the right panel of Figure \ref{fig:rs_ex}. We note here that we are neglecting any effects from potential ``recycling" of the core's atmosphere by the protoplanetary disk, but see, e.g. \cite{osk_2015} and \cite{ll_2017} for discussions of this effect. If the gas flow is able to penetrate into the core's atmosphere (e.g. \citealt{faw2015}), or if the core's atmospheric mass is small due to the high accretion luminosity, the core may be able to accrete the small particle sizes that we exclude. However, the accretion timescales for these particles are extremely long (see Section \ref{high_core}), so even if these particles can indeed accrete, their inclusion makes a negligible contribution to the total growth rate.

Cores that have $R_b < R$ will not be able to accrete a substantial amount of gas from the nebula, which occurs for planetary masses $M<M_a$, where
\begin{align}
M_a &\equiv \frac{c_s^3}{G}\left(\frac{3}{4 \pi G \rho_p}\right)^{1/2}\\
 &\approx 2 \times 10^{-4} M_\oplus \left( \frac{a}{30 \, \rm{AU}} \right)^{-9/14} \left( \frac{\rho_p}{2 \, \rm{g} \, \rm{cm}^{-3}} \right)^{-1/2}
\end{align}
where $\rho_p$ is the density of the protoplanet (e.g. \citealt{raf06}). The lowest core masses considered in this work are below this threshold. In this case, the considerations discussed above will still apply with the protoplanet's radius $R$ in place of its Bondi radius (i.e. accretion will cease for $R_{WS}^\prime < R$).

In summary, the largest sizes of particles that can deplete their kinetic energy can be captured at the core's Hill radius $R_H$. For smaller sizes of particles, the WISH radius will eventually become smaller than $R_H$, which limits the impact parameters where accretion can occur. Finally, the smallest sizes of particles will have $R_{WS}^\prime < R_b$. These particles will not be able to penetrate into distances $<R_{WS}^\prime$, and therefore will not be able to accrete via pebble accretion.

\subsection{Summary of Timescale Calculation} \label{t_grow}

In this section, we briefly discuss how $t_{\rm{grow}}$, as well as the parameters necessary for calculating the growth timescale ($R_{\rm{acc}}$, $H_{\rm{acc}}$, $H_p$, and $v_\infty$, see Equation \ref{eq:t_grow}), are determined. We also define symbols that will be used in the rest of the paper. For a summary of how these parameters are calculated, see the Appendix. For a more detailed discussion of how the calculation is performed, see R18. 

Besides the orbital separation, $a$, the mass of the planet, $M$, and the stellar mass, $M_*$, the other input parameters needed to calculate the growth timescale are the radius of the small bodies being accreted, $r_s$, and the strength of the turbulence, which is parameterized by the Shakura-Sunyaev $\alpha$ parameter (\citealt{ss_alpha}). 

Using the value of $r_s$, we can calculating the stopping time of the particle, $t_s$, and the particle's Stokes number,
\begin{align}
St \equiv t_s \Omega \; .
\end{align}
The Stokes number is a dimensionless measurement of the particle's size in terms of how well coupled the particle is to the gas and is the directly relevant parameter for calculating the effects of gas drag on the particle. We also note that using this form of the Stokes number for expressions involving turbulence (e.g. Equations \ref{turb_rel_gas} and \ref{eq:v_turb_kep}) implicitly assumes that the turnover time of the largest-scale turbulent eddies is equal to the local orbital period.

The value of $\alpha$ parameterizes the strength of the local turbulence in terms of the turbulent viscosity: $\nu_t = \alpha c_s H_g$, where $H_g = c_s/\Omega$ is the scale height of the gas disk. In terms of $\alpha$, the local turbulent gas velocity is given by
\begin{align} \label{eq:v_gas}
v_{\rm{gas},t} = \sqrt{\alpha} c_s \; .
\end{align}
We use $\alpha$ mainly to parameterize the magnitude of the turbulent gas velocity, which is the quantity that affects the pebble accretion process. It is only in Section \ref{final_mass} that we explicitly use $\alpha$ to parameterize the viscosity. While the $\alpha$ model of accretion disks is generally invoked to transport angular momentum inward and explain measured accretion rates in protoplanetary disks (see, e.g. \citealt{mlb_2005}), for our purposes, $\alpha$ is fundamentally a local parameter and is not necessarily connected with the accretion rate onto the central star. The most commonly cited mechanism for generating turbulence in protoplanetary disks is the magnetorotational instability (MRI; for a review, see \citealt{b_2009}). Simulations of MRI under ideal magnetohydrodynamical (MHD) conditions find effective $\alpha$ values of $10^{-2}-10^{-1}$ (e.g. \citealt{hgb_1995}), while MHD simulations that include nonideal MHD effects such as ambipolar diffusion find lower $\alpha$ values, in the range $10^{-4}-10^{-3}$ (e.g. \citealt{bs_2011}). In these simulations, the RMS turbulent gas velocity can be approximated to order of magnitude by taking $v_{gas} = \sqrt{\alpha} c_s$, as in Equation \eqref{eq:v_gas} (e.g. \citealt{xbmc_2017}). More recent works argue that magnetically driven winds can generate observed accretion rates, in which case protoplanetary disks could be quite inviscid (see, e.g. \citealt{b_2016}, \citealt{som_2016}). Even in this case, however, pure fluid instabilities, such as convective overstability (see, e.g. \citealt{l_2014}) or the zombie vortex instability (see, e.g. \citealt{mpj_2015}), spiral density waves raised by giant planets (see, e.g. \citealt{bnh_2016}), and hydrodynamical turbulence (see, e.g. \citealt{fnt_2017}), can all generate large RMS velocities for which the effective $\alpha$ value in Equation \eqref{eq:v_gas} is not equal to the $\alpha$ value characterizing angular momentum transport.

Once $a$, $M$, $M_*$, $r_s$, and $\alpha$ are specified, we can calculate the quantities needed to determine $t_{\rm{grow}}$. To begin, in order to determine the rate that particles encounter the core, as well as the kinetic energy of the small body relative to the protoplanet, we need to calculate the small body's velocity far from the core. Because we take the core to move at the local Keplerian velocity, we take $v_\infty$ to be set by the larger of the particle's shear velocity, $v_{\rm{shear}} = R_{\rm{acc}} \Omega$, and its velocity relative to the local Keplerian velocity, which is due to the particle's interaction with both the laminar and turbulent components of the gas velocity. We use $v_{pk}$ to denote the value of this velocity relative to the Keplerian orbital velocity. Thus, $v_\infty$ is given by
\begin{align}
v_\infty = \max(v_{pk},v_{\rm{shear}}) \; .
\end{align}

For every particle size, we calculate both the kinetic energy of the particle before the encounter, 
\begin{align} \label{eq:ke}
KE = \frac{1}{2} m v_\infty^2,
\end{align}
and the work done by gas drag during the encounter, 
\begin{align} \label{eq:work}
W =2 F_D(v_{\rm{enc}}) R_{\rm{acc}},
\end{align}
where $v_{\rm{enc}}$ is the velocity of the small body relative to the gas during its encounter with the core. For a discussion of how $F_D$ and $v_{\rm{enc}}$ are calculated, see the Appendix. Particles that have $KE > W$ cannot accrete; i.e. we set $t_{\rm{grow}} = \infty$ for such particles, regardless of the values of the parameters in Equation \eqref{eq:t_grow}. \footnote{Particles with $R_{\rm{acc}}=R_H$ and $v_\infty = v_H$ merely have their growth timescale enhanced by a factor $KE/W$ for $KE>W$, see the appendix for more details.} In practice, this sets the upper limit on the particle sizes that can be accreted via gas-assisted growth.

In addition to determining the work done on the particle during its encounter, the impact parameter for accretion $R_{\rm{acc}}$ is used to determine the width of the accretion cross section. As stated in Section \ref{part_size}, $R_{\rm{acc}}$ is given by \begin{align}
R_{\rm{acc}} = \min(R_{H}, R_{WS}^\prime) \; .
\end{align}
For more details on how $R_{WS}^\prime$ is calculated, see the Appendix. 

The height of the accretion rectangle $H_{\rm{acc}}$ is the minimum of the particle scale height $H_p$ and the impact parameter for accretion $R_{\rm{acc}}$:
\begin{align}
H_{\rm{acc}} = \min(R_{\rm{acc}}, H_{p}) \; ,
\end{align}
as particles with a vertical extent larger than $R_{\rm{acc}}$ will not be accreted. The particle scale height is also needed because it sets the density of the small bodies; it can be set by the Kelvin-Helmholtz shear instability or by turbulent diffusion,
\begin{align} \label{eq:h_p}
H_p =& \max(H_{KH},H_t) \nonumber \\ =& \max\left[\frac{2 \eta v_k}{\Omega} \min\left(1,St^{-1/2}\right),H_g\min\left(1,\sqrt{\frac{\alpha}{St}}\right) \right] \; ,
\end{align}
where $v_k$ is the local Keplerian orbital velocity, $\eta \equiv c_s^2/\left(2 v_k^2 \right)$ is a measure of the pressure support in the protoplanetary disk, and $\eta v_k$ is the velocity of the nebular gas relative to $v_k$ due to radial pressure support (i.e. the non-turbulent velocity of the gas). 

\subsection{Values of Parameters}

For the purposes of reporting numerical values in what follows, we use a fiducial set of parameters that specify the properties of the protoplanetary disk at a given semi-major axis. The effect of varying some of the parameters is discussed in R18.

We take the central star to be a solar mass, $M_* = M_\odot$. The small bodies and the core are taken to be spherical, with density $\rho_s = 2 \, \text{g} \, \text{cm}^{-3}$. We assume the gas disk is 70\% H$_2$ and 30 \% He by mass, leading to a mean molecular weight of $\mu \approx 2.35 \, m_H \approx 3.93\times 10^{-24} \, \text{g}$, with a neutral collision cross section $\sigma \approx 10^{-15} \, \text{cm}^2$. The temperature and gas surface density profiles are taken to be power laws in the semi-major axis. For the temperature profile we take $T = T_0 (a/\text{AU})^{-3/7} \, \text{K}$ (\citealt{cg_97}), where $T_0 = 200 \, \text{K}$, which is appropriate for a disk irradiated by star of luminosity $L \sim 3 L_\odot$ (e.g. \citealt{igm16}). For the gas surface density we use $\Sigma = 500 (a/\text{AU})^{-1} \, \text{g} \, \text{cm}^{-2}$,  and we assume a constant solid-to-gas mass ratio of $f_s = 1/100$. These choices are made to match solid surface densities found in observations of protoplanetary disks (see Figure \ref{fig:andrews09}).

\section{Gas-Assisted Growth Timescales at Wide Orbital Separation} \label{wide_sep}

In this section, we discuss the timescales for growth via pebble accretion at wide orbital separation ($\gtrsim 10$ AU), where canonical core accretion models are slow. We find that even in the presence of strong ($\alpha \gtrsim 10^{-2}$) turbulence, the growth timescales for high-mass cores are far shorter for pebble accretion than for planetesimal accretion. Indeed, the doubling timescale is so fast at these orbital separations that it begs the question of what inhibits this rapid growth. To investigate this question, we also show that, unlike for planetesimal accretion, gas-assisted growth is generally slower for low core masses, particularly when turbulence is strong. 

\subsection{Growth at High Core Mass} \label{high_core}

In the case of planetesimal accretion, the growth timescale for a core to reach $M_{\rm{crit}}$ is dominated by the last doubling timescale of the core; i.e. the slowest growth occurs for the highest core masses. Thus, when considering whether growth of wide orbital separation gas giants is possible, most authors examine the growth timescales at large core masses, which limit growth in canonical core accretion. 

The modification to the core accretion model presented by gas-assisted growth, on the other hand, substantially decreases the last doubling time at $M_{\rm{crit}}$ to well below the disk lifetime, even at wide orbital separations (e.g. \citealt{lj12}). While turbulence can reduce the rapid growth rates provided by pebble accretion, our modeling reveals that, at high core mass, growth remains efficient even in the presence of strong turbulence. This agrees with results from MHD simulations by \cite{xbmc_2017}, who numerically explore the growth rates of high-mass planetary cores in the presence of MRI turbulence.

\begin{figure} [h]
	\centering
\includegraphics[width=1.05\linewidth]{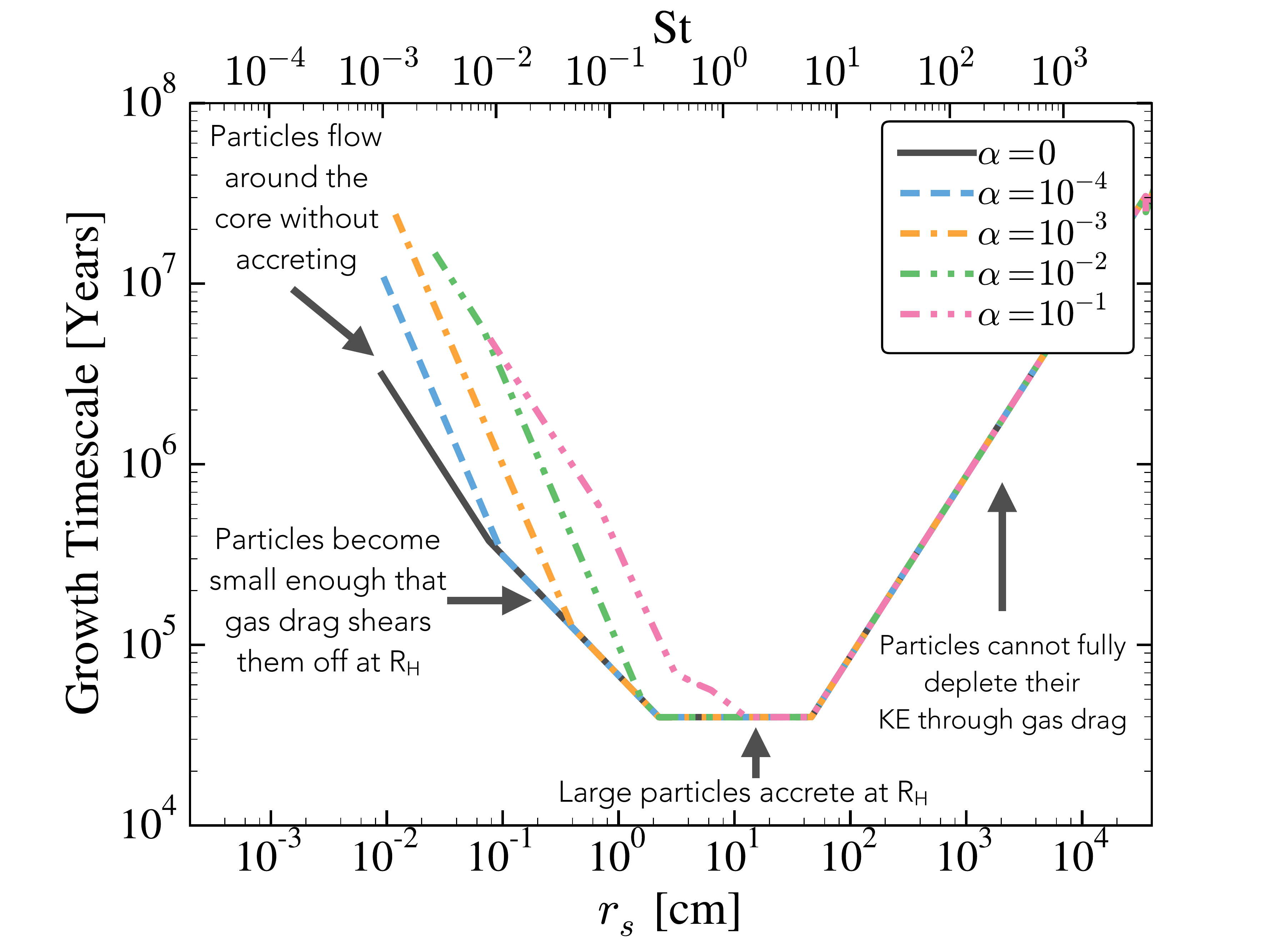}
	\caption{The growth timescale for a protoplanet as a function of the small-body radius the core is accreting. The timescale is plotted for several values of $\alpha$, which measures the strength of turbulence in the disk. The values shown are for a $5 \, M_\oplus$ core at $30 \, \text{AU}$. The lines are cut off for particles that are unable to accrete according to the energy criteria discussed in Section \ref{t_grow}.}
	\label{RvsT1}
\end{figure}

An example of our results is shown in Figure \ref{RvsT1}, which shows the growth timescale for a $5\,M_\oplus$ planet located at 30 AU. For particle sizes $r_s \gtrsim 50 \, \rm{cm}$, the growth timescale increases $\propto St$, as these particles require many orbital crossings to fully dissipate their kinetic energy (see the appendix and R18). As we decrease the small-body radius, we encounter small-body sizes $\left(1 \, \rm{cm} \lesssim r_s \lesssim 50 \, \rm{cm}\right)$ that are large enough that wind-shearing and scale height considerations are unimportant, allowing them to accrete over the entire Hill sphere at a rapid rate that is independent of $r_s$. As we continue to move to smaller pebble radii, eventually the particle size becomes small enough that the WISH radius and the particle scale height become important, decreasing the accretion rate. Finally, we reach the point where $R_{WS}^\prime < R_b$, which marks the pebble size at which particles couple so strongly to the gas that they flow around $R_{b}$ without accreting. This causes the cutoff in the graph seen on the left. For all values of $\alpha$ shown in Figure \ref{RvsT1}, there exists a broad range of small-body sizes, $r_s$, for which gas-assisted growth is able to operate, and the growth timescale of the core is less than the disk lifetime. Though turbulence erodes accretion of the smallest pebbles that were available in the laminar case, there still exists a range of particle sizes where rapid growth is possible.

Figure \ref{RvsT1} shows the emergence of a regime where the core can accrete larger particles at a minimal timescale that is independent of $r_s$ and $\alpha$. This timescale is reached for cores accreting in 2D (i.e. $H_{\rm{acc}} = H_p$) over the entirety of their Hill radius. As discussed in, e.g. R18, the maximal possible approach velocity in the 2D regime occurs when particles shear into $R_H$, i.e. when $v_\infty = v_H$; larger velocities will excite pebbles vertically, causing the core to accrete in 3D. Setting $H_{\rm{acc}} = H_p$, $R_{\rm{acc}} = R_H$ and $v_\infty = v_H$ in Equation \eqref{eq:t_grow}, we that see this timescale is given by
\begin{align} \label{eq:t_min}
t_{\rm{Hill}} = \frac{M}{2 f_s \Sigma R_H^2 \Omega} \; .
\end{align}
In terms of fiducial parameters, $t_{\rm{Hill}}$ can be expressed as
\begin{align} \label{eq:t_min_fid}
t_{\rm{Hill}} \approx 4 \times 10^4 \left( \frac{a}{30\,\text{AU}}\right)^{1/2} \left( \frac{M}{5\,M_\oplus} \right)^{1/3} \text{years}.
\end{align}

This timescale, which we will refer to as the ``Hill timescale," is faster than gravitational focusing by a factor $R_H^2/R_{\rm{focus}}^2 \approx R_H/R$. If we approximate the star and the planet as uniform density spheres and take $\rho_* \sim \rho_p$, we have $R_H/R \sim a/R_*$. Thus, not only is the enhancement in growth rate substantial, the enhancement of pebble accretion relative to gravitational focusing is an increasing function of semi-major axis.

\begin{figure*} [h]
	\centering
	\includegraphics[width=\linewidth]{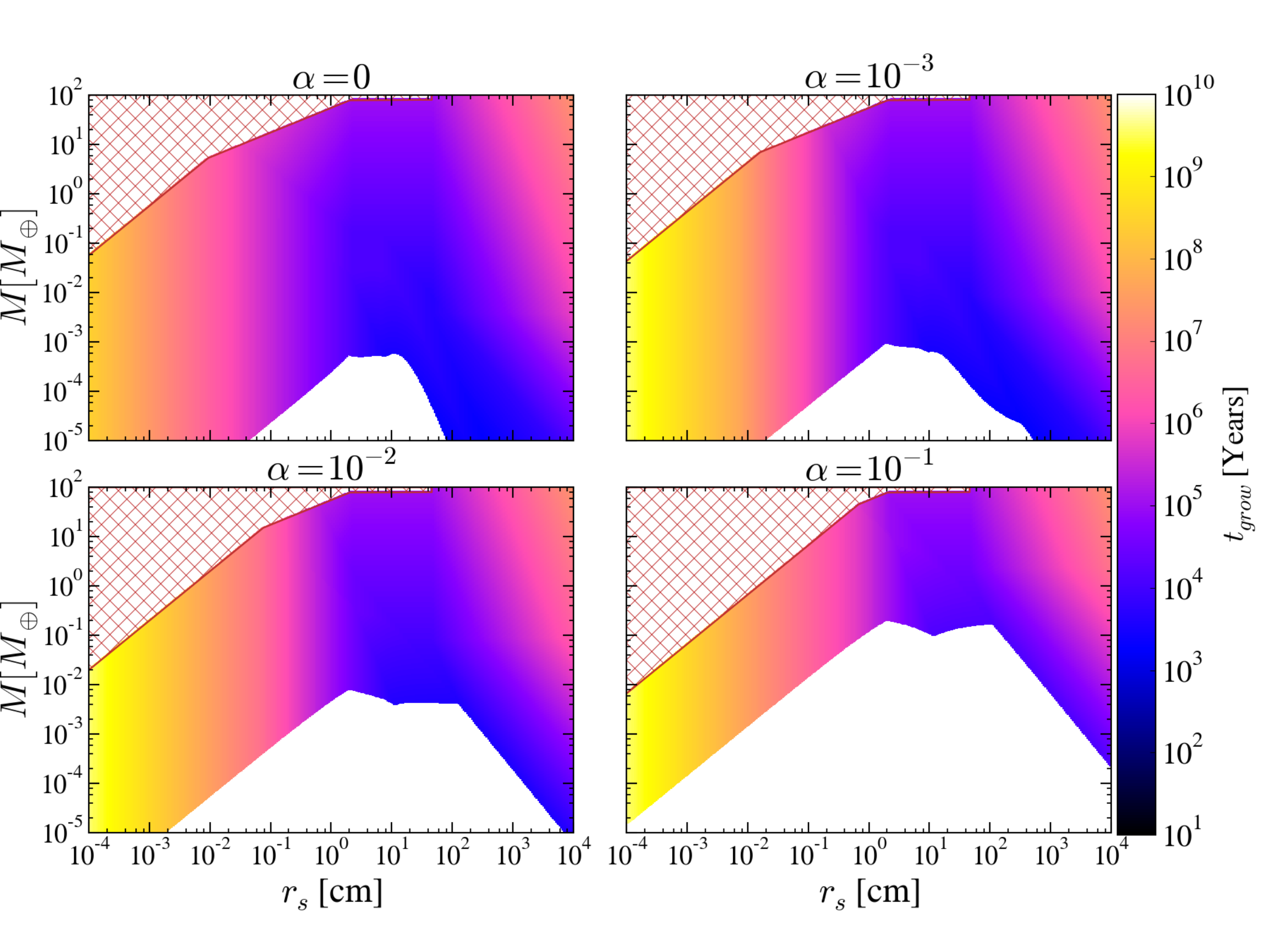}
	\caption{The growth timescale as a function of core mass, for $a = 30\,\text{AU}$. The red hatched region indicates where no accretion is possible. In the white regions particles can still accrete via other processes, e.g. gravitational focusing.}
	\label{fig:heatmap_a30}
\end{figure*}

The qualitative features of growth discussed above apply over a wide range of core masses. This can be seen from examination of Figure \ref{fig:heatmap_a30}, which shows the growth timescale for protoplanets as a function of both core mass and small-body radius. The four panels show the growth timescale for four different values of $\alpha$, while each individual panel shows the growth rate plotted as a function of both $r_s$ and $M$. As can be seen in Figure \ref{fig:heatmap_a30}, growth at ``high" core masses $\left( \gtrsim 10^{-3}-10^{-2} M_\oplus \right)$ proceeds in a similar manner to what is shown in Figure \ref{RvsT1}: the largest pebbles accrete on the rapid Hill timescale, independent of the small-body radius $r_s$, while smaller pebbles accrete less efficiently. Thus, as long as there exists a reservoir of particles that are able to accrete at $t_{\rm{Hill}}$, growth at higher core mass proceeds rapidly, even in the presence of strong turbulence.

However, it is also clear from examination of Figure \ref{fig:heatmap_a30} that below some ``minimum" mass, growth operates in a qualitatively different manner. We discuss the reasons for this change, as well as the ramifications for planetary growth, in the next section.

\subsection{Growth Timescales for Low-Mass Protoplanets}

In the last section, we showed that growth at large core masses is quite fast in gas-assisted growth, even in the presence of strong turbulence. This efficiency brings another issue, however, as we now need to understand why wide orbital separation gas giants are not ubiquitous given these rapid growth rates. As we will show below, at wide orbital separations and low core masses, growth timescales can be substantially longer than $t_{\rm{Hill}}$.

Figure \ref{fig:heatmap_a30} illustrates the difference in growth at low core masses. One feature of particular note in this figure is how the range of particle sizes available for accretion is restricted both at low core masses ($M \lesssim 10^{-3}-10^{-2} M_\oplus$ in the figure) and as the strength of turbulence increases. The limited range of sizes where pebble accretion can operate is often neglected in other works on pebble accretion, e.g. \cite{lj12}. While some works, such as \cite{OK10}, discuss an upper limit on particle size that agrees with our work in the laminar regime, this upper limit shrinks rapidly as the strength of turbulence is increased, as can be seen in the figure. For a more in-depth comparison of these models, see R18. 

In all four panels, we see that, for low core masses, it is primarily small particles with high accretion timescales that are available for growth. It is only when the core reaches a sufficiently ``large" mass \footnote{The mass scale where this change occurs is well approximated by $v_H/v_{\rm{gas}} = 48^{-1/3}$, see R18.} that the features discussed in the previous section emerge and cores are able to rapidly accrete pebbles. Thus, there is in some sense a ``minimum" mass, above which pebble accretion becomes efficient and proceeds on timescales less than the lifetime of the disk. This trend is more pronounced as the strength of turbulence is increased, in the sense that the mass required for accretion to be faster than the disk lifetime increases rapidly as $\alpha$ increases.

We note here that in the ``weak" turbulence regime (top panels of Figure \ref{fig:heatmap_a30}), there is a feature where the cores can accrete a limited range of sizes (with $r_s \sim $ 1 m) on short timescales. This is caused by the fact that the heaviest particles can actually have low kinetic energies relative to the core, since they drift at speeds close to the Keplerian velocity. This effect is eroded by the presence of turbulence, since it excites the random velocity of even the largest particles. In what follows, this effect is unimportant, since our choice of size distribution means that such large particles are not present (see Section \ref{int}), though it is an interesting area for future inquiry.  

The difficulty in accreting larger pebbles at low core mass is due to the weaker gravitational influence of the core. Gravitational perturbations from the core on the incoming pebbles can greatly increase the drag force on the small body during the encounter, since they increase the velocity of the particle relative to the local gas flow. The strength of this perturbation increases with increasing core mass. Thus, as core mass decreases, the work done by gas drag is reduced, limiting the range of small-body sizes that can be accreted. Furthermore, the size of $R_{\rm{acc}}$ also decreases with core mass, which means incoming particles have a smaller distance over which they can dissipate their kinetic energy relative to the core. Increasing the strength of the turbulence in the disk amplifies the difficulty in accreting particles, as incoming pebbles now have substantially higher kinetic energies. 

Not only can a smaller range of particles be accreted at low core mass, the smaller particle sizes that are available for growth have long growth timescales. One reason for this is that smaller particles can be more easily pulled off the core by gas drag, meaning that their maximum impact parameter for accretion is $R_{\rm{acc}} = R_{WS}^\prime$, which can be quite small in comparison to $R_H$. Furthermore, these particles are more easily excited vertically by the turbulent gas velocity (see Equation \ref{eq:h_p}). Thus, smaller particles have larger scale heights, reducing their number density and further slowing growth. 

Thus, at lower core masses, we expect growth via pebble accretion in high turbulence to be quite slow, as the only particles that the smaller cores can accrete have large growth timescales. These timescales can be several orders of magnitude slower than the growth timescale at $M_{\rm{crit}}$, meaning that growth at low core masses can often be the time-limiting step in gas giant formation via gas-assisted growth.

\subsection{Size Distribution of Small Bodies} \label{int}
Because the growth timescale is generally much slower for smaller particle sizes than it is for larger ones, the timescale for growth will also be dependent on the size distribution of small bodies that are available for accretion. Thus, in order to facilitate a more quantitative discussion of where growth of gas giants is possible in our model, we will integrate quantities of interest over an assumed size distribution, a process we now discuss in more detail. 

If the size distribution of small bodies is specified, i.e. if we know $d N/dr_s$, we can integrate the accretion rate of the large body over small body radius and obtain a total accretion rate. This integrated timescale is sensitive to the actual form of size distribution employed; thus, while integrating over size distribution can be quite illustrative, the results are less general.

For our purposes, we employ the power-law distribution from \cite{size_dist}, who calculated the steady-state size distribution from a collisional cascade. This gives a distribution of sizes such that $dN/dr_s \propto r_s^{-3.5}$. For a power-law size distribution $dN/dr_s \propto r^{-q}$, most of the mass is in the largest particle sizes for $q<4$. Thus, our results are insensitive to the lower cutoff radius but highly dependent on the upper radius, since for an $r_s^{-3.5}$ power-law, most of the mass is in the larger particles. This is the most important feature of the size distribution we employ: for any size distribution with most of the mass in the largest particle radii, the qualitative picture discussed below is unchanged, though the quantitative results will change by order unity factors.

Unless otherwise stated, we use an upper radius such that the largest Stokes number present is $St_{\rm{max}} = 10^{-1}$, and a lower radius such that the smallest bodies present correspond to $St_{\rm{min}} = 10^{-4}$. This constant Stokes number upper limit is most appropriate for the case when particle growth is limited by collisions, and the relative velocity is dominated by laminar drift. The value of $St_{\rm{max}} = 10^{-1}$ comes from \cite{blum_wurm_coll}, who gave $r_s = 10$ cm as the size past which collisions become destructive for a particular set of disk parameters. This radius corresponds to $St \sim 10^{-1}$ for the disk they considered. If bodies are held together mainly by chemical forces, then the relative velocity between particles is the main determinant of the outcome of a collision. This relative velocity in turn depends on the particle Stokes number and the amplitude of the gas velocity. Because the laminar drift velocity $\eta v_k$ is approximately constant throughout the disk, if laminar drift sets the collision velocity, the particle Stokes number is the only parameter relevant to determining when collisions become destructive.

This simple description of the size distribution neglects the effects of increased turbulence, which would increase the particle-particle relative velocities during a collision, in turn lowering the critical Stokes number for destructive collisions. This also neglects the importance of radial drift in the outer regions of protoplanetary disks, which can proceed on shorter timescales than particle-particle collisions. In general, the size distribution of pebbles in disks is more complex than the simple prescription given here. We use this as our fiducial size distribution in order to reduce the number of input parameters our results depend on while still describing the general features of gas-assisted growth.

\subsection{Integrated Growth Timescales}

In this section, we discuss how integrated growth timescales change as the core grows. We also discuss how we can analytically calculate the integrated growth timescale, which is used later on to calculate analytic expressions for both the minimum mass for pebble accretion to be rapid and the semi-major axes where gas giant growth can occur. 

\begin{figure} [h]
	\centering
	\includegraphics[trim={5cm 0 0 0},clip,width=1.15\linewidth]{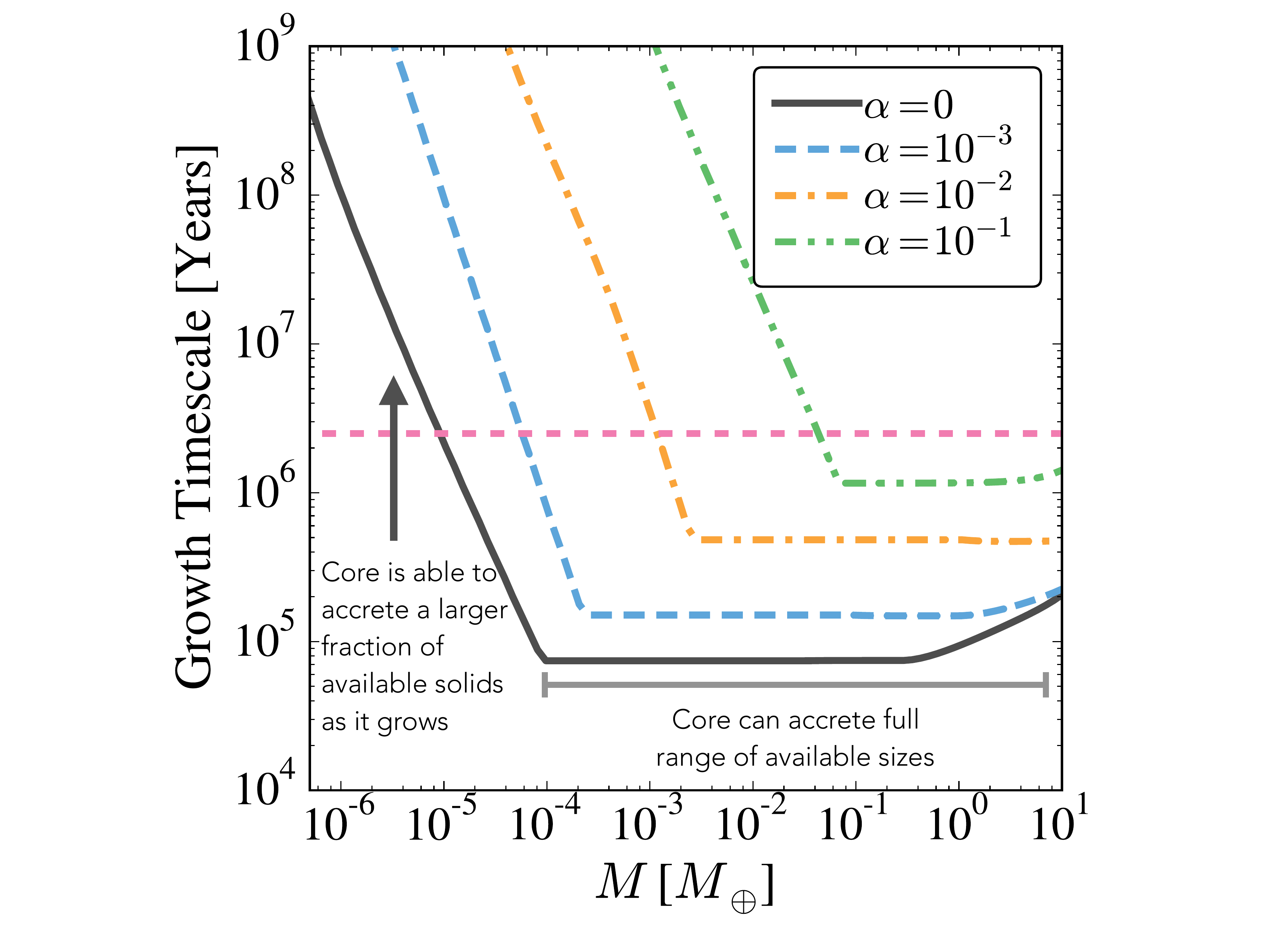}
	\caption{Integrated growth timescale for a core at $a = 20 \, \rm{AU}$ as a function of core mass. The growth timescale is integrated over sizes using a Dohnanyi distribution with a maximum size corresponding to $St=10^{-1}$, as discussed in Section \ref{int}. The approximate $e$-folding time of the gaseous component of the disk, $\tau_{\rm{disk}}$, is marked as a dashed horizontal line. As the core grows, it can accrete a larger fraction of the available small-body sizes, causing the growth timescale to drop rapidly. Eventually, the core's mass becomes large enough that it can accrete all available particle sizes, causing it to enter into a regime where growth timescale is independent of $M$. We also note that once the core becomes massive enough that the growth timescale drops below $\tau_{\rm{disk}}$, subsequent growth at higher core masses proceeds on timescales well below the disk lifetime.}
	\label{fig:t_vs_m}
\end{figure}

An example of the results from integrating over small-body size is shown in Figure \ref{fig:t_vs_m}, which plots the integrated growth timescale at $a = 20 \, \text{AU}$ as a function of core mass for several different levels of turbulence. An estimate for the $e$-folding time for the dissipation of the gaseous component of the disk, $\tau_{\rm{disk}} \approx 2.5 \, \text{Myr}$, is also shown. The disk dissipation timescale $\tau_{\rm{disk}}$ represents an approximate cutoff for gas giant formation; cores that are unable to reach the critical core mass within $\tau_{\rm{disk}}$ will not be able to trigger runaway accretion before the gas is substantially depleted. 

At low core masses, the growth timescale drops quickly as the core mass increases. This is due to the fact that these larger cores can accrete more massive pebbles. For the lower-mass cores, the largest pebbles that the core can accrete are smaller than the maximal size of the particles present. Therefore, as the core grows, it can accrete a larger fraction of the available solids, increasing the growth rate. 

If the growth timescale has only a simple power-law dependence on $r_s$ for the whole range of sizes, we can explicitly integrate the growth timescale over size and calculate an analytic expression for $t_{\rm{grow}}$. This requires that none of the parameters that go into calculating $t_{\rm{grow}}$ change regimes over the range of sizes considered: for example, if $R_{\rm{acc}} = R_H$ for the largest sizes present but $R_{\rm{acc}} = R_{WS}^\prime$ for the smaller sizes, our integrand is now a piecewise function of $r_s$, and a simple analytic solution is no longer possible. In practice, if we make the approximation that the regimes that apply for the maximal particle size present hold throughout the integral, the resultant errors are generally small. 

In what follows, we will be particularly interested in the mass at which the growth timescale becomes shorter than the lifetime of the gas disk, since subsequent growth will proceed on even shorter timescales. From Figure \ref{fig:t_vs_m}, we can see that, at the point where $t_{\rm{grow}}$ becomes shorter than $\tau_{\rm{disk}}$, cores are small enough that $R_{\rm{acc}} = R_{WS}^\prime$ (i.e. the core's WISH radius is smaller than its Hill radius for all of the small-body sizes it accretes)\footnote{As discussed in the appendix, $R_{WS}^\prime$ is really the smaller of two radii: $R_{WS}^\prime = \min\left(R_{WS},R_{\rm{shear}}\right)$. In making our analytic approximations, we assume that the cores we are concerned with are low enough mass that $R_{WS}<R_{\rm{shear}}$. This assumption can be shown to be generally valid by comparing the analytic approximations we derive to our numerical results}. We also assume that $R_{WS}^\prime$ is small enough that the core accretes in 3D, i.e. $R_{WS}^\prime < H_p$. Additionally, for the small-body radii the core is accreting $v_\infty = v_{pk} \approx v_{\rm{gas}}$ (i.e. the small body's random velocity dominates over shear, and these particles are well coupled to the gas). Due to the wide orbital separations and small particle sizes we are interested in, the particles are expected to be in the Epstein drag regime. Using these values throughout the integration over size allows us to compute $t_{\rm{grow}}$ analytically, and comparison of the resultant analytic expressions with the numerical calculations presented below shows that these approximations are robust. Finally, when calculating the work done on the particle, we set $v_{\rm{enc}} = v_{\rm{kick}} = G M / \left( R_{\rm{acc}} v_\infty \right)$ (see the appendix for more details). It can be shown that particles with the Stokes number given by Equation \eqref{eq:st_limit} are in this regime.

Using the considerations above, we can now calculate closed-form expressions for $t_{\rm{grow}}$. To begin, we determine the largest size of particle these low-mass cores can accrete. Using Equations \eqref{eq:t_s}, \eqref{eq:ke}, and \eqref{eq:work} and the values of parameters discussed in the preceding paragraph, we see that the maximal size of particle the core can accrete is given by\footnote{
	We note that this is similar to the barrier between the ``Hyperbolic" and ``Full Settling" regimes identified by \cite{OK10}, except that our value for $v_{\rm{gas}}$ includes a contribution from the turbulent gas velocity, which dominates for $\alpha > \eta$. 
	}
\begin{align} \label{eq:st_limit}
St_\ell = 12 \frac{v_H^3}{v_{\rm{gas}}^3} \; .
\end{align}

Because our size distribution is dominated by the largest particle sizes, we can use the Stokes number limit from Equation \eqref{eq:st_limit} to determine the growth rate. If we neglect the lower bounds on integrations over particle size, it is straightforward to demonstrate that the growth rate is, to order-of-magnitude, given by the product of the growth rate for the largest sizes of particles the core can accrete, $\dot{M}(St_\ell)$, and the fraction of the surface density contained in solids up to size $St_\ell$, $f(St_\ell)$:
\begin{align}
\dot{M} \sim \dot{M}(St_\ell) f(St_\ell) \; .
\end{align}

Plugging in our assumed values for the parameters into Equation \eqref{eq:t_grow} for $t_{\rm{grow}}$, we see that in this regime,
\begin{align}
t_{\rm{grow}} = \frac{H_p}{2 \Sigma_p G t_s} \; .
\end{align}
Thus, there are two possible growth regimes, depending on whether $H_p = H_{KH}$ or $H_p = H_t$. For $St < 1$, we have $H_t > H_{KH}$ (i.e. $H_p = H_t$) for $\alpha > 2 \eta St$. This limit on $St$ divides our analytic expressions into two piecewise regimes.

Explicitly performing the integration over size, the growth timescale is given by

\begin{align} \label{eq:frac_an}
t_{\rm{grow}} \approx \begin{dcases}
 9 \times 10^{7} \, \text{years} \, \left( \frac{M}{10^{-5} M_\oplus} \right)^{-2} \times \\ \quad \left( \frac{a}{30 \, \, \text{AU}} \right)^{5/2} St_{\rm{max}}^{1/2} \left( \frac{\alpha}{10^{-3}} \right)^{7/2} ,& \alpha > 2 \eta St_\ell \\ \\
 3 \times 10^{7} \, \text{years} \, \left( \frac{M}{10^{-5} M_\oplus} \right)^{-3/2} \times \\ \quad \left( \frac{a}{30 \, \text{AU}} \right)^{51/14} St_{\rm{max}}^{1/2} \quad, & \alpha < 2 \eta St_\ell
\end{dcases}
\end{align}

Eventually, the core becomes massive enough that it can accrete all sizes of particles available, i.e. $St_{\ell} > St_{\rm{max}}$. This causes the growth timescale to become independent of $M$, since $\dot{M} \propto R_{WS}^{\prime 2} \propto M$. In this regime, the growth timescale is given by
\begin{align} \label{eq:full_an}
t_{\rm{grow}} \approx \begin{dcases}
7 \times 10^{3} \, \text{years} \, \left( \frac{a}{30 \, \text{AU}} \right)^{11/14} \times \\ \quad St_{\rm{max}}^{-3/2} \left( \frac{\alpha}{10^{-3}} \right)^{1/2}, & \alpha > 2 \eta St_{\rm{max}}\\ \\
1 \times 10^{4} \, \text{years} \, \left( \frac{a}{30 \, \text{AU}} \right)^{15/14} St_{\rm{max}}^{-1}, &\alpha < 2 \eta St_{\rm{max}}
\end{dcases}
\end{align}

Thus, the low-mass growth of the core, where the growth timescale decreases for increasing core mass, is the time-limiting step in gas-assisted growth. As can be seen in Figure \ref{fig:t_vs_m} and verified by the analytic expressions above, once the core reaches a mass such that $t_{\rm{grow}} < \tau_{\rm{disk}}$, all subsequent growth should proceed on timescales that are faster than the disk lifetime. Therefore, the early stages of core growth, where gravitational focusing of planetesimals may be faster than gas-assisted growth, will play a key role in whether a planet can grow to be a gas giant.

\section{Restrictions on the Growth of Gas Giants} \label{gas_giants}

The effects of the previous sections imply that to understand under what conditions gas giant formation is possible via pebble accretion, we must examine lower-mass cores, for which the gas-assisted growth timescale can be quite long. If these cores were to grow by gas-assisted growth alone, then growth would always stall at sufficiently small core mass such that turbulence dominates over the core's gravity. For low core masses, however, planetesimal accretion can be quite rapid. Therefore, the final fate of a protoplanet depends on whether canonical core accretion can provide sufficiently rapid growth at small core masses such that the core can reach a size where pebble accretion becomes efficient, which will in turn allow the core to grow rapidly to the critical core mass needed for runaway growth.

\subsection{Planetesimal Accretion Timescale }
In order to calculate the semi-major axis where gas giants form, we consider early growth by planetesimal accretion and subsequent growth by pebble accretion. This requires us to calculate the timescale for growth by planetesimal accretion for a given core mass. In general, the scale height of particles is given by $H_p = v_z/\Omega$, where $v_z$ is the vertical component of the small body's velocity. As stated previously, the fastest growth possible via planetesimal accretion (without some external damping mechanism) occurs when the planetesimal velocity dispersion is equal to the Hill velocity. We use this regime for our fiducial value of the growth timescale via planetesimal accretion. If we take $v_z \sim v_\infty =v_H$, and use Equations \eqref{eq:m_dot} and \eqref{eq:r_focus}, then the growth rate of the core is proportional to

\begin{align}
\dot{M} \propto R_H  R \, \Sigma_{\rm{pla}} \Omega 
\; ,
\end{align}
where $\Sigma_{\rm{pla}}$ is the surface density of planetesimals. The  prefactor in the above equation is not well constrained by analytic considerations; in a more detailed treatment, it should taken from $N$-body simulations of the interactions between the planetesimals. For our purposes, we take the prefactor from \cite{jl_17}; this gives

\begin{align}
\dot{M} = 6 \pi R_H  R \, \Sigma_{\rm{pla}} \Omega \; .
\end{align}
For our fiducial value of the growth timescale, we set $\Sigma_{\rm{pla}} = \Sigma_p$, which gives a timescale of 
\begin{align} \label{eqn:t_GF_fid}
t_{\rm{pla}} \approx 2 \times 10^{7} \, \text{years} \left( \frac{a}{30 \, \text{AU}} \right)^{3/2} \left( \frac{M}{5 \, M_\oplus} \right)^{1/3}.
\end{align} 
Solving for the mass where $t_{\rm{pla}}=\tau_{\rm{disk}}$ gives an expression for the maximum mass a planet can reach via planetesimal accretion, 
\begin{align}
M_{\rm{pla}} = 8 \times 10^{-3} M_\oplus \fid{a}{30 \, \rm{AU}}{-9/2} \fid{\Sigma_{\rm{pla},0}}{5 \, \rm{g} \, \rm{cm}^{-2}}{3} \; ,
\end{align}
where $\Sigma_{\rm{pla},0}$ is the prefactor of the planetesimal surface density profile, i.e. $\Sigma_{\rm{pla}} = \Sigma_{\rm{pla},0} \left(a/\rm{AU}\right)^{-1}$. Our choice of $\Sigma_{\rm{pla}} = \Sigma_p$ gives reasonable values of the masses planets can reach within the gas disk lifetime at the semi-major axes of the solar system gas giants (see Figure \ref{fig:m_min}). Some of the effects of varying the surface density of planetesimals are discussed in Section \ref{upper_lim}.

\subsection{Upper Limits on the Semi-Major axis of Gas Giant Growth} \label{upper_lim}

In order to place constraints on the semi-major axis at which gas giant growth is possible, we begin by determining the minimal mass below which pebble accretion is too slow to grow a core within $\tau_{\rm{disk}}$. In order to do this, we make the approximation that once the core becomes massive enough that $t_{\rm{grow}}<\tau_{\rm{disk}}$, the growth timescale of the core will remain below $\tau_{\rm{disk}}$ as the core continues to grow to $M_{\rm{crit}}$. Thus, once the core becomes massive enough, its subsequent growth time is small compared to the disk lifetime. As can be seen in Figure \ref{fig:t_vs_m}, and from Equations \eqref{eq:frac_an} and \eqref{eq:full_an}, this approximation is quite robust. An exploration of $t_{\rm{grow}}$ vs. $M$ over a large amount of parameter space shows that this is generally true throughout the disk. We note, however, that if the sizes of the available particles are not set by Stokes number but rather by absolute particle size, there can exist regions of the disk where this approximation breaks down, as the particle sizes where growth is efficient are essentially determined by Stokes number.  We consider this possibility in more detail below. 

\begin{figure} [h]
	\centering
	\includegraphics[width=\linewidth]{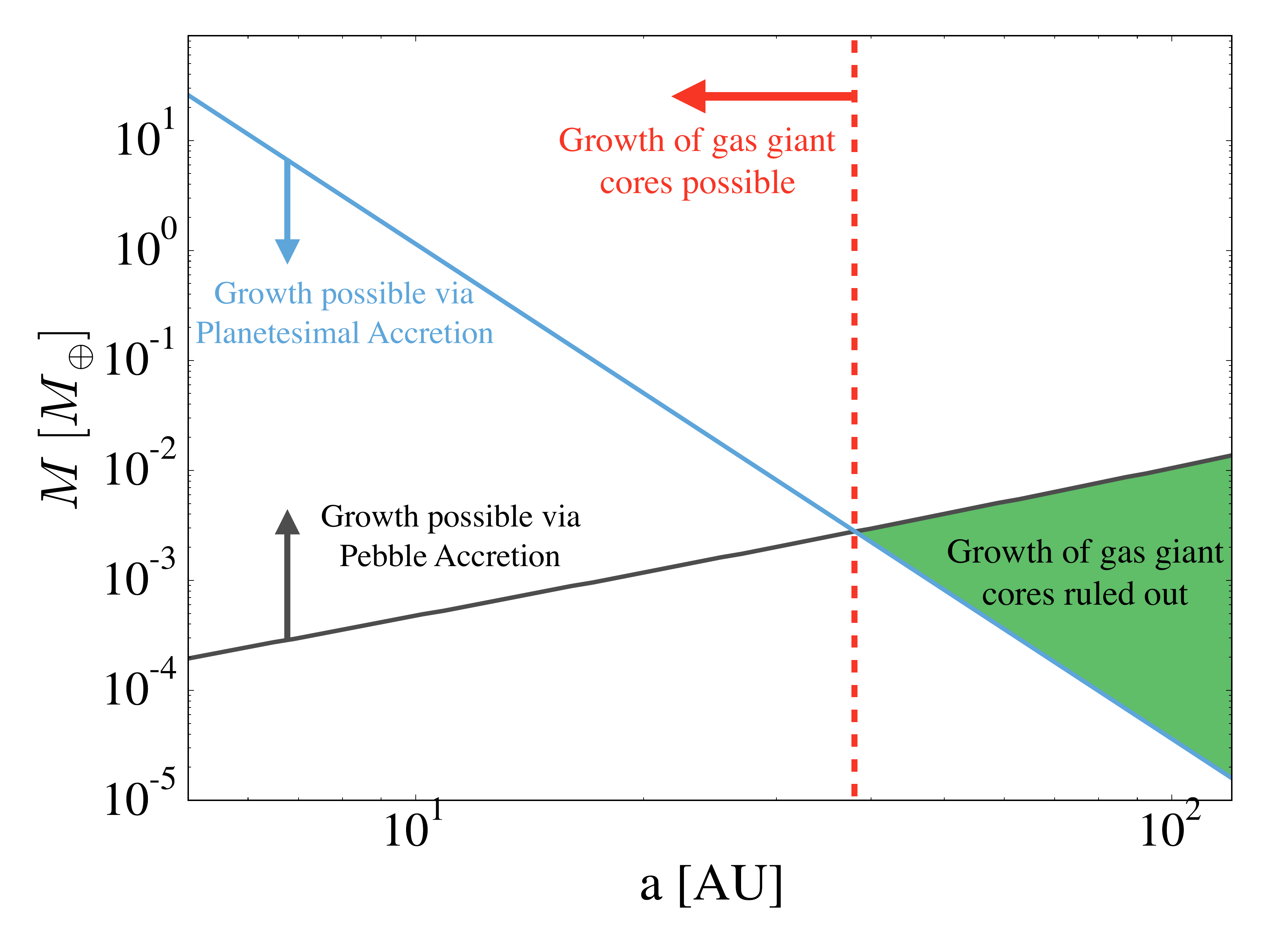}
	\caption{Graphical illustration of how gas giant core growth is limited for different semi major axes. The monotonically increasing line (black) shows the \textit{minimum} mass needed for gas-assisted growth to produce a gas giant core; for masses higher than the plotted mass, the growth timescale for the core is less than the disk lifetime. The monotonically decreasing line (blue) shows the \textit{maximum} mass it is possible to achieve via planetesimal accretion. Values lower than the indicated mass can be reached within the disk lifetime, but for larger masses the disk will dissipate before the mass is reached. The vertical line denotes the semi-major axis upper limit on where growth of gas giant cores can occur; interior to this region, planetesimal accretion can build a massive enough core rapidly enough that pebble accretion becomes efficient and dominates growth at higher masses. The green shaded region indicates where growth of gas giants is ruled out, as both planetesimal accretion and pebble accretion are too slow.}
	\label{fig:m_min_ex}
\end{figure}

Because the growth timescale is dominated by growth at low core masses, we can determine an approximate minimum mass for gas giant growth through pebble accretion by solving for the mass at which $t_{\rm{grow}}=\tau_{\rm{disk}}$. This is the mass below which growth will stall, and the core will be unable to grow to $M_{\rm{crit}}$ within $\tau_{\rm{disk}}$. This idea is shown graphically in Figure \ref{fig:m_min_ex}.

\begin{figure} [h]
	\centering
	\includegraphics[width=\linewidth]{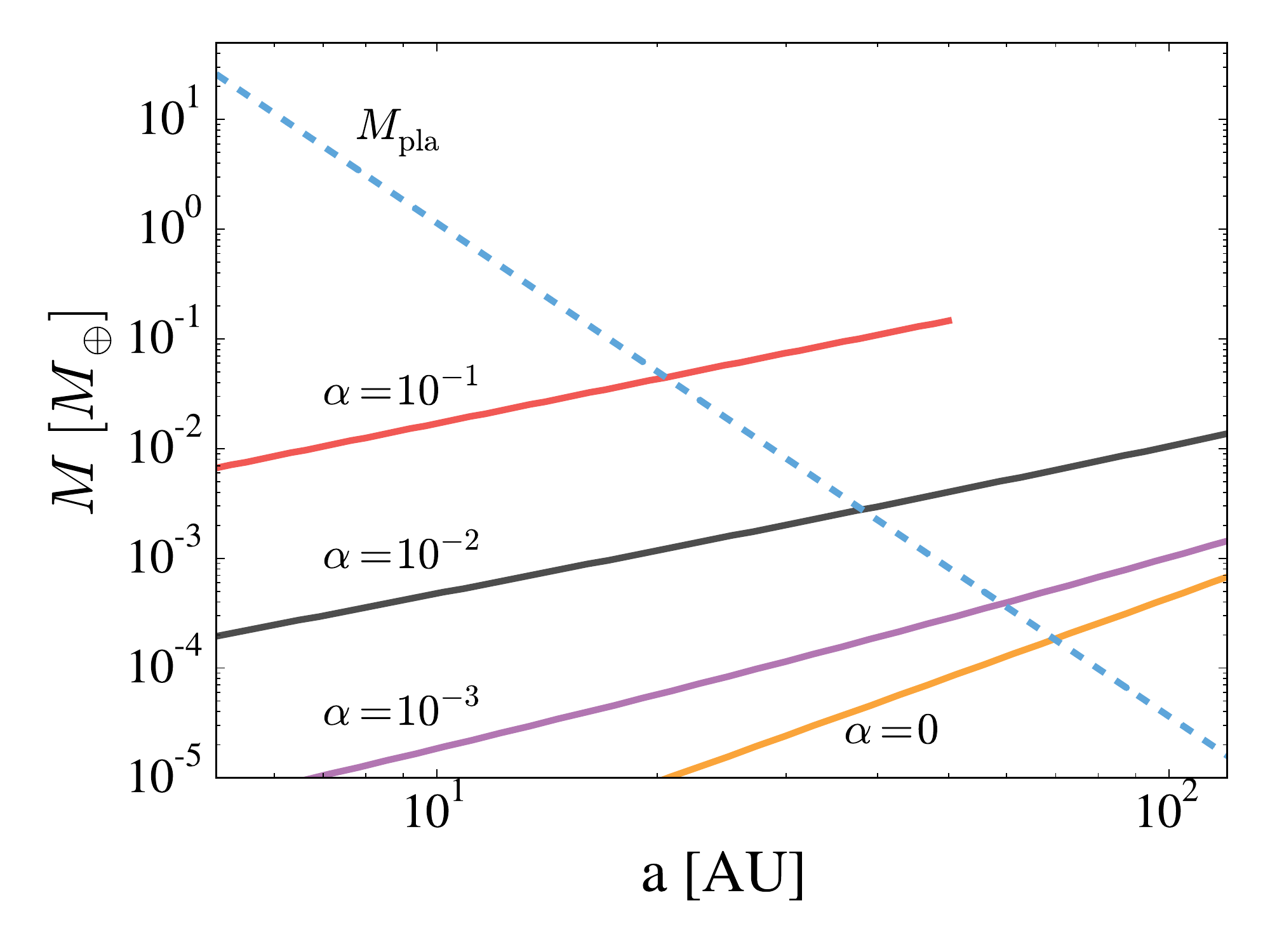}
	\caption{Minimum mass for which the growth timescale is shorter than $\tau_{\rm{disk}} = 2.5 \, \text{Myr}$, shown for various values of $\alpha$. Masses smaller than the values depicted have growth timescales larger than $\tau_{\rm{disk}}$, so if the core can exceed this mass by other means, it should be able to reach $M_{\rm{crit}}$, but growth by pebble accretion will be unable to exceed this mass within $\tau_{\rm{disk}}$. A mass for growth by planetesimal accretion is also shown, but this mass has a different interpretation: it is the \textit{largest} mass a core can grow to via gravitational focusing within $\tau_{\rm{disk}}$.}
	\label{fig:m_min}
\end{figure}

The mass where growth stalls as a function of semi-major axis, calculated numerically using our full expressions, is shown in Figure \ref{fig:m_min}. Again, the effects of turbulence on growth rate are clearly visible in the figure: in the laminar case, even extremely wide orbital separation cores can grow faster than $\tau_{\rm{disk}}$ down to a very low core mass. At high turbulence ($\alpha \gtrsim 10^{-2}$), however, the core needs to reach masses $\gtrsim 10^{-3} M_\oplus$ before pebble accretion becomes fast enough for these cores to reach $M_{\rm{crit}}$ within the disk lifetime. Also shown in the plot is the \textit{maximum} mass that a core can grow to using gravitational focusing of planetesimals. We emphasize here that the interpretation of this line is the opposite of the gas-assisted growth values; for gravitational focusing, all values \textit{lower} than the given mass are approximately obtainable within the disk lifetime.

We can also obtain an analytic approximation for $M_{\rm{peb}}$, the mass where the pebble accretion timescale drops below the disk lifetime. Setting \eqref{eq:frac_an} equal to $\tau_{\rm{disk}}$ gives
\begin{align} \label{eq:m_peb}
M_{\rm{peb}} = \begin{dcases}
6 \times 10^{-5} \, M_\oplus \left( \frac{a}{30 \, \text{AU}} \right)^{5/4}  \left( \frac{\alpha}{10^{-3}} \right)^{7/4} \times\\ \quad \left( \frac{\tau_{\rm{disk}}}{2.5 \, \text{Myr}} \right)^{-1/2}  St_{\rm{max}}^{1/4}, \quad \alpha > 2 \eta St_{\rm{max}}\\
5.6 \times 10^{-5} \, M_\oplus \left( \frac{a}{30 \, \text{AU}} \right)^{17/7} \times\\ \quad \left( \frac{\tau_{\rm{disk}}}{2.5 \, \text{Myr}} \right)^{-2/3}   St_{\rm{max}}^{1/3}, \quad \alpha < 2 \eta St_{\rm{max}}
\end{dcases}
\end{align}
which demonstrates analytically the strong dependence that the efficiency of pebble accretion has on both semi-major axis and strength of turbulence.

Figure \ref{fig:m_min_ex} shows how we can use Figure \ref{fig:m_min} to determine where the interplay between canonical core accretion and gas-assisted growth will allow a gas giant to grow.  The intersection between the pebble accretion and planetesimal accretion values represents the approximate semi-major axis upper limit on gas giant growth. For values higher than this semi-major axis, planetesimal accretion is too slow to bring the core to the minimum mass needed such that gas-assisted growth can subsequently grow the core to $M_{\rm{crit}}$ within the disk lifetime. For values smaller than this semi-major axis, however, planetesimal accretion can grow the core to a sufficiently massive size rapidly enough that gas-assisted growth can take over. This semi-major axis also represents an upper limit on where a core can form, as for smaller orbital separations, the growth timescale decreases (see Equations \ref{eq:frac_an} and \ref{eq:full_an}). This is not the case if the size distribution is determined by particle radius instead of Stokes number, as we discuss in Section \ref{fix_rs}. We also note that if a core larger than the pebble accretion mass were present past this semi-major axis limit (e.g. if it were scattered outward), then the core could grow sufficiently rapidly to trigger gas giant formation.

\begin{figure} [h]
	\centering
	\includegraphics[width=\linewidth]{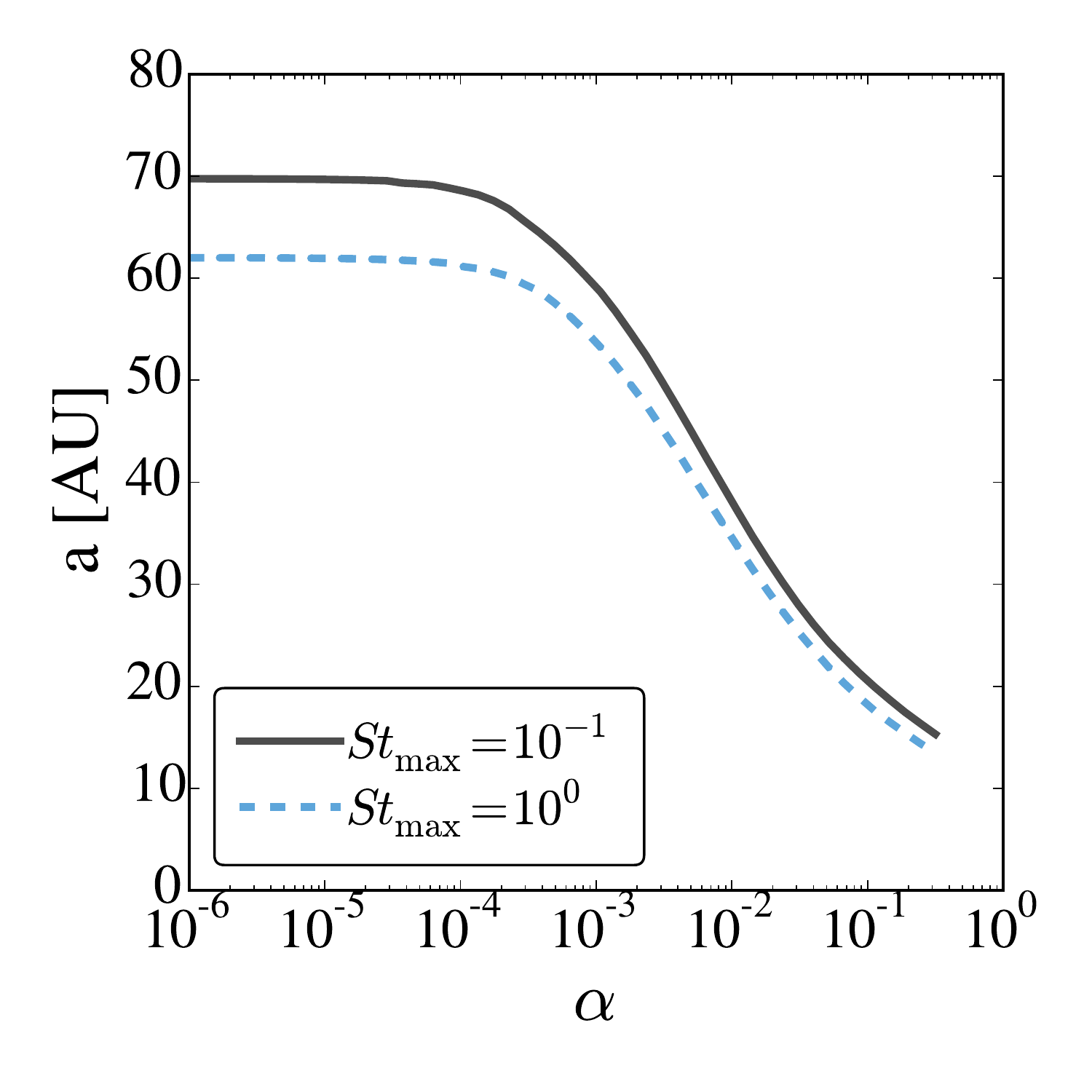}
	\caption{Maximum semi-major axis at which growth to critical core mass is possible as a function of $\alpha$. Curves are shown for a Dohnanyi distribution with a maximum-sized particle corresponding to $St=10^{-1}$ (solid line) and $St=1$ (dashed line). }
	\label{fig:a_max_alph}
\end{figure}

Figure \ref{fig:a_max_alph} plots the maximum distance obtained by solving numerically for the mass at which $M_{\rm{pla} } = M_{ \rm{peb}}$ using our full expressions. In order to illustrate the effect of changing the upper limit on the size distribution, two different size distributions are shown -- one in which the maximum Stokes number is $St_{\rm{max}} = 0.1$, and one in which $St_{\rm{max}} =1$.  From the plot, it is clear that as turbulence increases, the semi-major axis at which gas giant growth is possible drops substantially. Growth is also slightly more inhibited for the $St=1$ distribution; this is due to the fact that cores need to reach higher masses in order to accrete $St=1$ particles as opposed to $St=0.1$ particles. However, because $St=1$ particles accrete more rapidly, this effect is attenuated, causing the overall dependence on $St_{\rm{max}}$ to be rather weak.  

Using Equation \eqref{eq:m_peb}, we can derive analytic approximations to the curve shown in Figure \ref{fig:a_max_alph}. Setting $M_{\rm{peb}} = M_{\rm{pla}}$, we obtain,
\begin{align} \label{eq:a_max}
a_{\rm{upper}} \approx \begin{dcases}
70 \, \text{AU} \left( \frac{\alpha}{10^{-3}} \right)^{-7/23} \left( \frac{\Sigma_{\rm{pla}}}{5 \, \text{g} \, \text{cm}^{-2}} \right)^{12/23} \times \\ \left( \frac{\tau_{\rm{disk}}}{2.5 \, \text{Myr}} \right)^{14/23} St_{\rm{max}}^{-1/23}, \quad \quad \, \, \, \alpha > 2 \eta St_{\rm{max}}\\  
60 \, \text{AU}  \left( \frac{\Sigma_{\rm{pla}}}{5 \, \text{g} \, \text{cm}^{-2}} \right)^{42/97} \times \\ \left( \frac{\tau_{\rm{disk}}}{2.5 \, \text{Myr}} \right)^{154/291} St_{\rm{max}}^{-14/291}, \quad \alpha < 2 \eta St_{\rm{max}}
\end{dcases} 
\end{align}

\begin{figure} [h]
	\centering
	\includegraphics[width=1.0\linewidth]{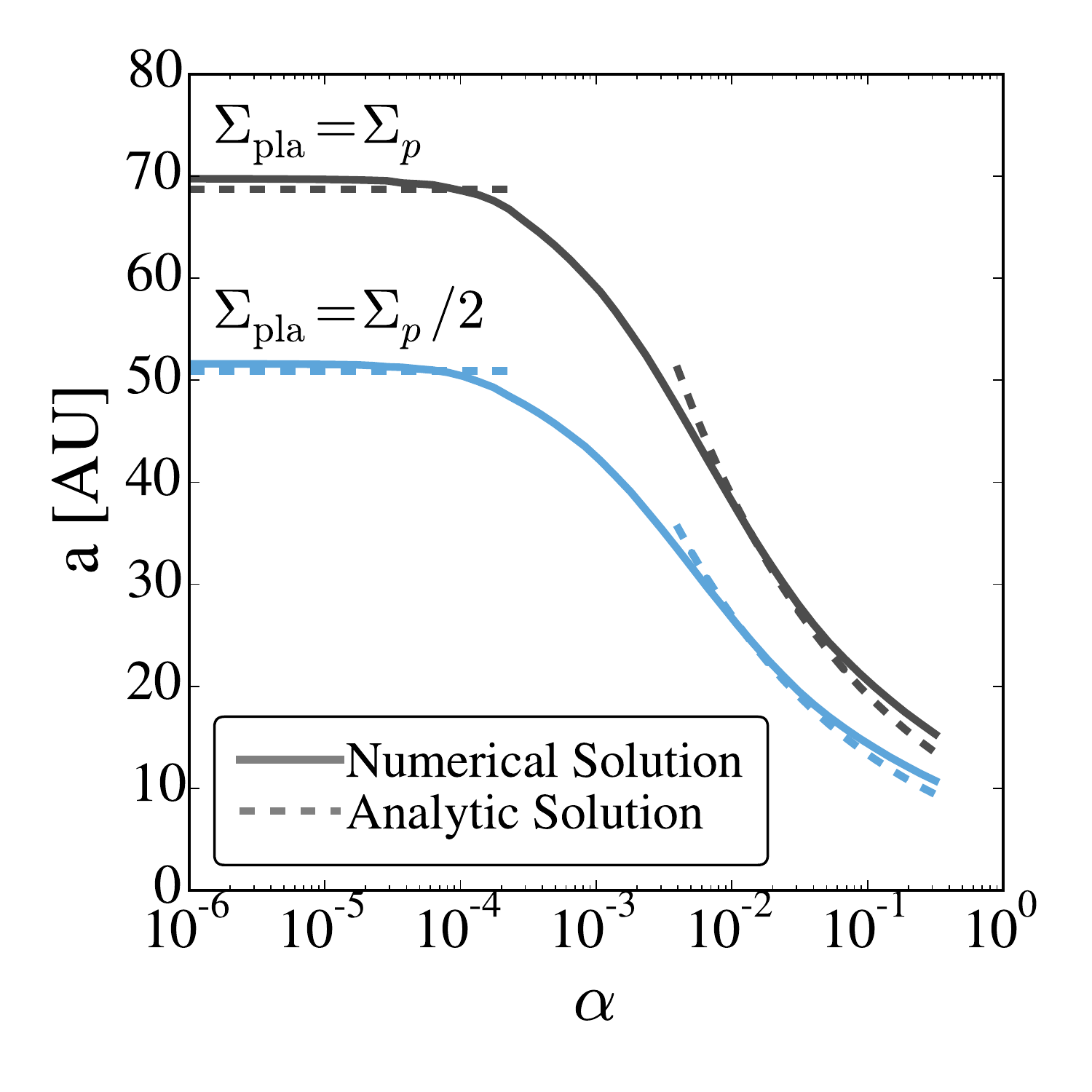}
	\caption{A comparison of our analytic expression for the maximal semi-major axis where gas giant growth is possible (Equation \ref{eq:a_max}), with the numerical solution. Results are presented for two different planetesimal surface densities, $\Sigma_{\rm{pla}}=\Sigma_p$ and $\Sigma_{\rm{pla}}=\Sigma_p/2$. }
	\label{fig:red_pla}
\end{figure}

These analytic expressions are overplotted on the numerical results in Figure \ref{fig:red_pla}. Curves for two different planetesimal surface densities, one where we use our fiducial value of $\Sigma_{\rm{pla}} = \Sigma_p$ and one where we have reduced the surface density by a factor of 2, are shown. As expected, our analytic results agree well with the full numerical calculation in the limits of small and large $\alpha$. Figure \ref{fig:red_pla} also demonstrates that reducing the planetesimal surface density can have a marked effect on the semi-major axis where gas giant growth is possible. 

The considerations discussed above can provide a plausible mechanism by which the growth of gas giants is suppressed: higher values of turbulence inhibit core growth at lower masses and make it so rapid growth via pebble accretion can only proceed once higher values of mass are reached. Because planetesimal accretion is slow at these wide orbital separations, cores will stall in their growth at low mass and be unable to reach the high masses needed for gas giant growth to proceed. In our order-of-magnitude model, the actual values quoted are of less import than the scalings and overall behavior predicted by the model. Thus, while we would not expect the quoted limit of e.g. $a \lesssim 30 \text{AU}$ for gas giant growth at $\alpha \approx 10^{-2}$, to be precise, we would argue that we should expect the general inhibiting of pebble accretion, and therefore gas giant growth, for stronger values of turbulence.

\subsection{Effect of Fixing Upper Particle Radius} \label{fix_rs}

Thus far, we have fixed the upper limit of our size distribution in terms of particle Stokes number. In contrast, disk models that are used to fit to observations of protoplanetary disks tend to use size distributions with fixed maximum particle radius instead of Stokes number. Size distributions fixed by particle radius can also emerge naturally if drift limits particle size as opposed to collisions. For example, \cite{pms_2017} derived an expression for the gas surface density determined by particle drift, which can be rewritten as an expression for particle radius (see their Equation 8):
\begin{align}
r_s = \frac{\Sigma a  }{t_{\rm{disk}} \eta v_k \rho_s} \; ,
\end{align}
where $t_{\rm{disk}}$ is the age of the disk. If $\Sigma_g \propto a^{-1}$, then the only semi-major axis dependence in the above equation comes from $\eta v_k$, which has extremely shallow radial dependence (e.g. $\eta v_k \propto a^{1/14}$ for the temperature profile we employ). Therefore, we also present results that use a size distribution where the upper size limit is fixed by particle radius. We follow the disk models of \cite{andrews_09}, who used a Dohnanyi ($d N/ds \propto r_s^{-3.5}$) distribution, with $r_{s,\rm{min}} = 0.005 \, \mu \text{m}$ and $r_{s,\rm{max}}  = 1 \, \text{mm}$. This 1 mm maximum size is consistent with fitting of disk spectral energy distributions (\citealt{dal_disk_models}).

\begin{figure} [h]
	\centering
	\includegraphics[width=\linewidth]{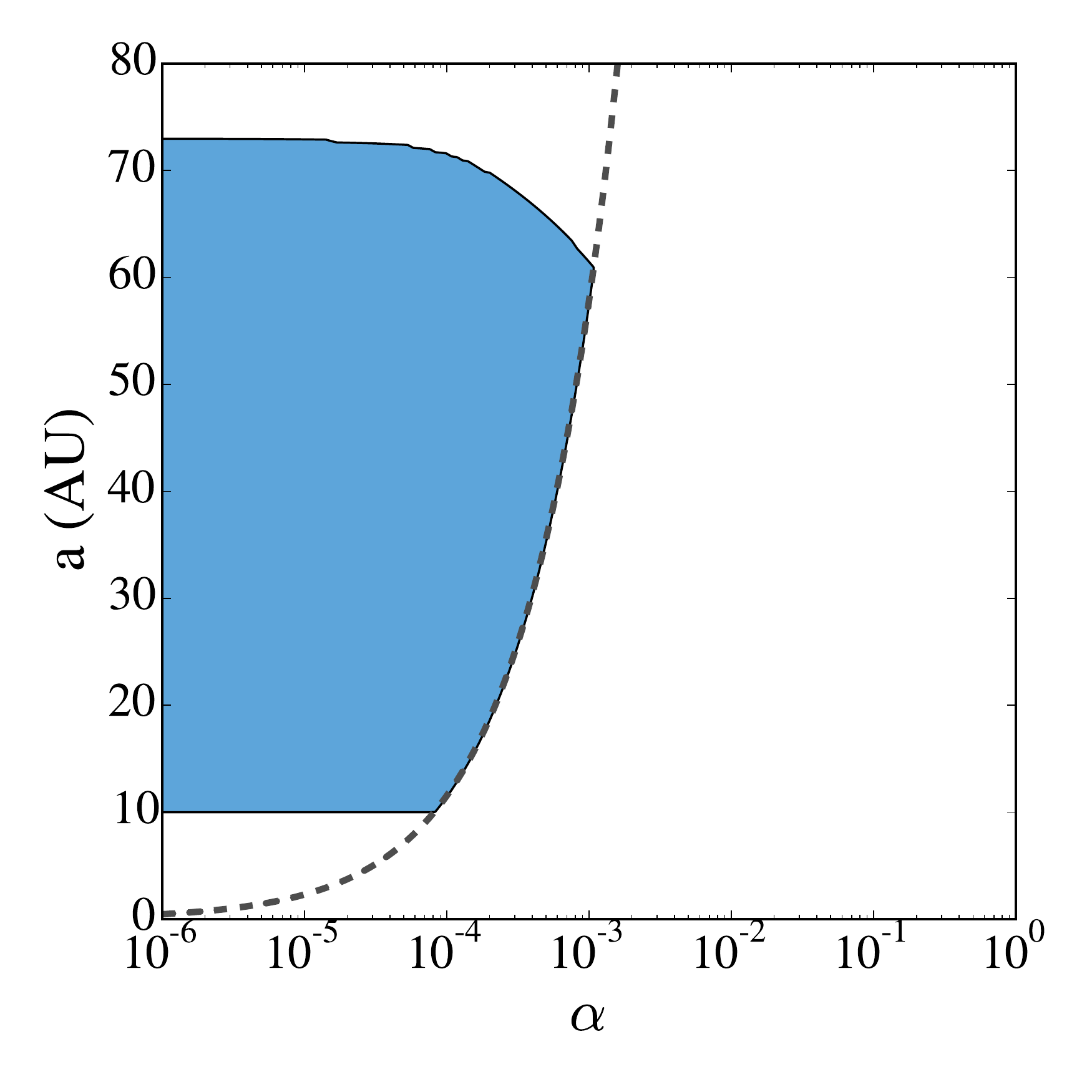}
	\caption{The blue region shows where growth of gas giant cores is possible for a size distribution with maximal pebble size of $r_s = 1\, \text{mm}$, plotted as a function of the strength of turbulence. In contrast to the size distributions which used a fixed Stokes number as the upper limit, this distribution has a lower limit on where core growth can occur as well as an upper limit. The lower limit is the maximum of a fixed semi-major axis limit, and a limit for a given $\alpha$ -- an analytic expression for the latter (c.f. Equation \ref{eq:a_low}) is also shown (black dashed line).}
	\label{fig:a_max_alph_1mm}
\end{figure}

A plot of the numerical solution for the semi-major axes where gas giant growth is possible for a distribution with fixed size limits is shown in Figure \ref{fig:a_max_alph_1mm}. The blue region indicates where growth is possible. As can be seen from the figure, using $r_s = 1\,\text{mm}$ as an upper size limit throughout the disk has pronounced effects on the semi major axes available for gas giant growth. For $\alpha \gtrsim 10^{-4}$, the region where gas giants can grow shrinks rapidly, causing a complete cutoff in gas giant growth for $\alpha \gtrsim 10^{-3}$. In this regime, we therefore expect turbulence to completely inhibit gas giant growth, instead of restricting growth to smaller values of semi-major axis.

Using an upper limit fixed by particle size leads to a lower limit on semi-major axis in addition to an upper limit. This lower limit stems from the fact that fixing a maximum particle radius means that even the largest sizes of particles present may have low Stokes numbers, causing them to be accreted inefficiently or not accreted at all. This introduces two additional processes that we need to consider when calculating where gas giants can form, one that gives a fixed semi-major axis limit independent of $\alpha$, and another that gives a lower limit on $a$ for a given $\alpha$.  

Firstly, cores accreting the maximum size of particle may not be able to grow to $M_{\rm{crit}}$ and trigger runaway gas accretion, as the Bondi radius may grow larger than $R_{WS}^\prime$ for the maximal particle size (and therefore for all smaller values of $r_s$) before $M =  M_{\rm{crit}}$. This means that all available particle sizes will be in the regime where they flow around the core's atmosphere without being accreted (see the right panel of Figure \ref{fig:rs_ex}), which will halt growth via pebble accretion. A core will have $R_{WS}^\prime = R_b$ when it reaches a mass of
\begin{align}
M_{R_{WS}^\prime=R_b} \approx 10 M_\oplus \left( \frac{a}{10 \, \text{AU}} \right)^{11/7} \left( \frac{r_s}{1 \, \text{mm}} \right) \; ,
\end{align}
where we have assumed the particle is in the Epstein drag regime in converting from $t_s$ to $r_s$. Since the mass where this equality occurs is an increasing function of semi-major axis, these considerations imply that, close in to the star, the core may not be able to reach sufficient mass through pebble accretion to trigger runaway gas accretion. Using $M_{\rm{crit}} = 10 M_\oplus$ as a conservative upper limit for runaway accretion to occur requires $a \gtrsim 10 \, \text{AU}$ before cores can reach $M_{\rm{crit}}$. 

A second complication that can also serve to place a lower limit on semi-major axis is that, unlike for the fixed Stokes number size distribution, growth timescale can be a \textit{decreasing} function of semi-major axis when particle radii are instead fixed. In particular, the mass-independent growth timescales for accretion of the full range of sizes given by Equation \eqref{eq:full_an} can decrease as we move outwards in the disk. This stems from the fact that the Stokes number of $r_s =$ 1 mm particles will increase further out in the disk, and particles with higher values of $St$ (for $St < 1$) are generally accreted more rapidly. Thus, even if we find an upper limit on semi-major axis in the manner described above, we also have to check whether the growth timescale again becomes longer than the disk lifetime closer in to the central star. Because the low-$\alpha$ regime shown in Figure \ref{fig:a_max_alph_1mm} occurs for small values of semi-major axis, where the Stokes number of an $r_s=1\,\rm{mm}$ particle is quite low ($\sim 5\times10^{-4}$), the $ \alpha >2\eta St_{max}$ regime of Equation \eqref{eq:full_an} applies everywhere when calculating our semi-major axis lower limit. We can therefore determine our lower limit on growth analytically by setting this timescale equal to $\tau_{\rm{disk}}$. Doing so yields
\begin{align} \label{eq:a_low}
a_{\rm{low}} \approx 58 \, \text{AU} \left( \frac{\alpha}{10^{-3}} \right)^{7/10}  \left( \frac{r_{s,\text{max}}}{1 \, \text{mm} } \right)^{-21/10} \left( \frac{\tau_{\rm{disk}}}{2.5 \, \text{Myr}} \right)^{-7/5} \; ,
\end{align}
which is plotted in Figure \ref{fig:a_max_alph_1mm} (black dashed line).

These two processes are what yields the lower limit seen in Figure \ref{fig:a_max_alph_1mm}; regardless of the value of $\alpha$, core growth cannot proceed for $a \lesssim 10 \, \rm{AU}$, as the core will be isolated from accretion of all available pebble sizes before it can trigger runaway gas accretion. As $\alpha$ increases, the growth timescale for accretion of the full range of particle sizes may become longer than the disk lifetime close in to the central star, requiring the core to be at larger values of semi-major axes before gas giant growth is possible.

\section{Final Mass of Gas Giants} \label{final_mass}

In this section, we consider the effect of $\alpha$ on the final mass that gas giants can reach and tie these considerations to our previous discussions on how turbulence affects the early stages of gas giant growth. Note that in this section, we take the $\alpha$ value to affect the viscosity of the disk, as opposed to merely using $\alpha$ to parameterize the RMS gas velocity, as was done in the previous sections.

Once a core begins runaway gas accretion, the accretion rate for nebular gas is initially extremely rapid (e.g. \citealt{pollack_gas_giants}). If accretion proceeded unhindered at this rate, gas giants would easily be able to accrete all of the gas in their local feeding zones before the gas disk dissipated. However, the observed masses of gas giants are well below their local gas isolation mass; what, then, stops gas giants from growing? This is usually explained by appealing to gap opening by the growing planet: as the planet grows, it will gravitationally torque the local nebular gas, pushing it away. If gas is torqued away more rapidly than it is transported inward by viscosity, the planet can clear a gap in the disk, reducing the gas surface density near the growing planet. This reduction in surface density can starve the planet of material for growth and can eventually shut off growth entirely. If this process sets the final mass that gas giants can reach, then, in general, gas giants will be able to reach larger masses in disks that have higher viscosities. Thus, if we translate our $\alpha$ values into viscosities (as opposed to just parameterizations of the local turbulent gas velocity), then turbulence can play a role in \textit{both} whether a gas giant core can form and in the final mass of the planet. 

The physical processes that determine the final mass of gas giants remain an open question meriting further inquiry. In order to provide concrete numerical results, in what follows, we consider two possible criteria from the literature for determining the mass that gas giants reach, and we discuss the implications for the population of wide orbital separation gas giants when these criteria are coupled with growth via pebble accretion. Thus, while the expressions we use may not capture the final masses of gas giants, the results below will still hold qualitatively as long as disks with higher viscosity produce higher-mass gas giants.

For our first criterion from the literature, we determine the width of the gap opened by the planet and shut off accretion when the gap has reached a certain size. The width of the gap opened can be obtained by equating the rate of angular momentum transport due to viscosity, $\dot{H_\nu} = 3 \pi \Sigma \nu a^2 \Omega$, with the rate that the planet delivers angular momentum to the disk (\citealt{lin_gap}):

\begin{align}
\dot{H}_T = f_g q^2 \Sigma a^4 \Omega^2 \left(\frac{a}{\Delta}\right)^3.
\end{align}
Here $q\equiv M/M_*$ is the planet-to-star mass ratio, $\Delta$ is the width of the gap opened, and $f_g$ is an order unity factor. Equating these two expressions gives the gap width in units of the Hill radius as

\begin{align}
\frac{\Delta}{R_H} = \left(\frac{f_g q}{\pi \alpha} \frac{a^2}{ H_g^2}\right)^{1/3}
\end{align}
From comparison with the results of numerical simulations of the growth of Jupiter by \cite{lhd_2009}, \cite{kratter_gas_giants} adopted $\Delta/R_H \sim 5$ as their criterion for starvation, which we adopt as well. Scaled to fiducial parameters, \citeauthor{kratter_gas_giants} gave the starvation mass as

\begin{align} \label{eq:m_starve}
M_{\text{starve}} \approx 8 M_J \left( \frac{\alpha}{4 \times 10^{-4}} \right) \left( \frac{T}{40 \, \text{K}} \right) \left( \frac{a}{70 \text{AU}} \right) \left( \frac{\Delta}{5 R_H} \right)^3 \; .
\end{align}

\begin{figure} [h]
	\centering
	\includegraphics[width=1.0\linewidth]{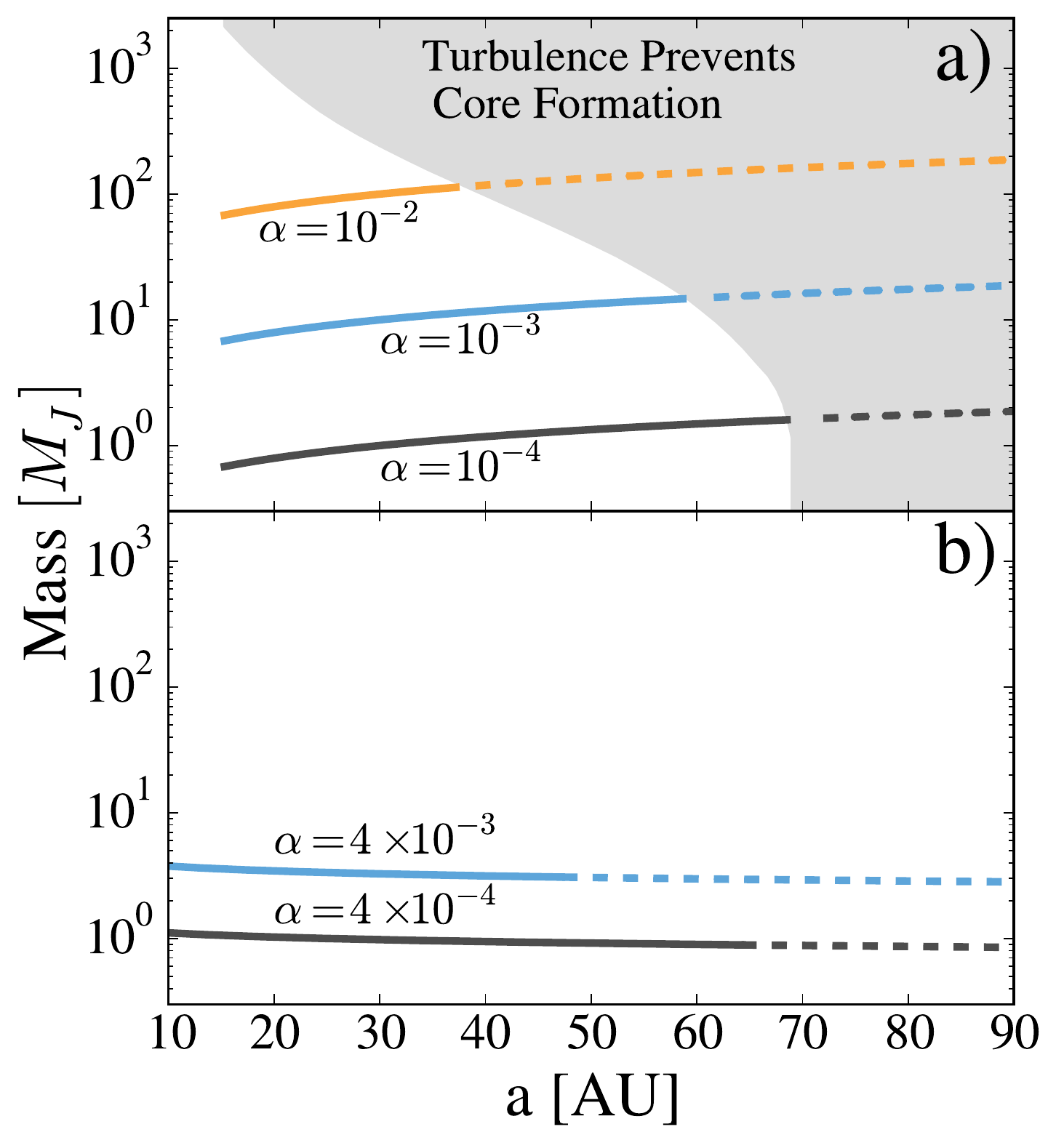}
	\caption{``Starvation" mass, past which growth of a gas giant halts, plotted as a function of semi-major axis. \textit{Panel a):} In this panel, the mass is obtained by assuming that growth shuts off when the planet opens up a gap of width $\Delta = 5 R_H$ (see text for more details). Inside the gray region, the $\alpha$ value required to prevent gap opening before the core reaches the given mass is so large that a core will not be able to form within the lifetime of the gas disk (see Section \ref{upper_lim}). The shape of this region is determined using our upper limits on $\alpha$ taken from the $St_{\rm{max}} = 0.1$ in Figure \ref{fig:a_max_alph}. The labeled curves show maximum masses for constant values of $\alpha$. \textit{Panel b):} Here the starvation mass is determined numerically using fitting formulae to numerical results for the gas accretion rate from 3D hydrodynamical simulations by \cite{lhd_2009}. The starvation mass is determined by solving for the mass at which $M_{\rm{starve}}/\dot{M} = \tau_{\rm{disk}}$. The dashed lines again indicate the semi-major axes where turbulence prevents a gas giant core from forming via pebble accretion.}
	\label{fig:m_starve}
\end{figure}

Using this expression for the starvation mass, we can calculate the final mass of gas giant planets in our model as a function of the strength of turbulence in the disk. An example is shown in the upper panel of Figure \ref{fig:m_starve}.

The gray region shows limits on gas giant mass, which are obtained by using our values for the $St_{\rm{max}} = 0.1$ curve in Figure \ref{fig:a_max_alph}. For points inside the gray region, in order to grow a gas giant up to the given mass, the viscosity needs to be so large that early stages of growth are too slow for a core to reach $M_{\rm{crit}}$ within the lifetime of the gas disk. Said another way, for semi-major axes and masses inside the gray region, growth of gas giants is ruled out using the criteria described in Section \ref{upper_lim}. Also plotted in Figure \ref{fig:m_starve} are the upper mass limits for several constant $\alpha$ values. When these curves enter the gray region, growth of gas giants is ruled out in our model. For low levels of turbulence, growth of gas giants can proceed out to large semi-major axes, but the final masses of these planets are low. As turbulence increases, opening a gap in the disk becomes harder, allowing the gas giant planets to reach higher masses, but the semi-major axes at which growth can occur become more restricted. We stress that this is a general feature of gas giant growth via pebble accretion and, in particular, is independent of the criterion used to determine the final mass of gas giants. Because viscous torques oppose the torque from the growing planet, the final mass the gas giant reaches will increase with the viscosity in the disk. Thus, if growth of gas giants proceeds by pebble accretion, we expect disks with higher viscosities to host more massive gas giants at smaller orbital separations.

While the torques from the growing planet can increase the width of the gap as the planet grows, material can still flow through the gap opened by the planet (e.g. \citealt{fsc_2014}). For example, \cite{msc_2014} showed that meridional circulation can still transport material from the top layer of the disk, which may imply that consideration of gap opening alone is insufficient to determine the final mass of gas giants. Thus, we present an alternate criterion for gap opening that takes into account gas accretion rates taken directly from the 3D hydrodynamical simulations performed by \cite{lhd_2009}. \citeauthor{lhd_2009} provided numerical results for the upper limit on the planet's gas accretion rate as a function of planetary mass for $\alpha = 4 \times 10^{-3}$ and $4 \times 10^{-4}$. Using these accretion rates, we can determine the mass past which $t_{\rm{grow}} = M/\dot{M} > \tau_{\rm{disk}}$. This scale represents another way of determining the starvation mass, since planets larger than this value will not grow substantially before the nebular gas dissipates. In practice, we can use the fitting formula given in Equation (2) of \cite{lhd_2009} for $\alpha = 4 \times 10^{-3}$ to determine this mass numerically as a function of semi-major axis. For $\alpha = 4 \times 10^{-4}$ the authors did not provide a closed-form expression; we instead interpolate between the values plotted in their Figure 2 to obtain $\dot{M}$ as a a function of $M$. 

The result of this calculation is shown in the lower panel of Figure \ref{fig:m_starve}. As can be seen from the figure, this criterion leads to lower starvation masses in comparison with halting growth at a fixed value of gap width. As in the upper panel, the dashed sections of the lines indicate where the timescale to form the core exceeds the lifetime of the gas disk. Unlike the upper panel however, it is not possible to indicate the overall region where this occurs, as \cite{lhd_2009} did explicitly calculate $\dot{M}$ as a function of $\alpha$.

These considerations could provide an explanation for the proposed correlation between stellar mass and gas giant frequency. While the dynamical changes due to increasing stellar mass have a relatively minor effect on core growth rates (R18), these higher-mass stars are expected to have substantially higher luminosities with moderately higher amounts of ionizing radiation  (e.g., \citealt{pf_2005}, \citealt{wmw_2010}). Disks with higher ionization fractions should have higher levels of MHD turbulence (e.g. \citealt{a_2011}), leading to higher effective $\alpha$ values. Thus, from the above considerations, we should expect that higher-mass stars will yield higher-mass planets. Furthermore, if disk mass is correlated with stellar mass (\citealt{pth_2016}) and the final mass of gas giants is set by accretion rate (as opposed to gap width, where the disk surface density cancels out), then we would expect planets around more massive stars to accrete at higher rates and therefore reach higher masses before their growth timescale becomes lower than the disk lifetime.  Thus, the gas giant planets found around more massive stars may represent the high-mass tail of a distribution of gas giants formed at large distances via gas-assisted growth, which have their final mass dictated by gap-opening criteria. If this is the case, then we would expect that there exists a population of gas giants around these stars as well that are simply lower mass than can be detected with the current generation of imaging instruments. This is easily accomplished if these planets are $\lesssim 2 M_J$; see, e.g. \cite{bowler_DI_review}.

\section{Summary/Conclusions} \label{summary}

In this paper, we have used our previously discussed model of gas-assisted growth in a turbulent disk to study the problem of growth of gas giants at wide orbital separations. At these large distances, last doubling timescales for growth by planetesimal accretion are far longer than the disk dispersal timescale of the gas, making growth of gas giants by canonical core accretion extremely difficult.

Gas-assisted growth allows cores to easily complete their last doubling time to critical core mass, even in strong turbulence. The maximal growth rate provided by pebble accretion, $t_{\rm{Hill}}$, is extremely rapid, even in the outer disk. For massive cores. even strong turbulence does not substantially inhibit growth.

The same is not true for smaller core masses. however. Growth of gas giants at large distances can easily stall at smaller core masses. By integrating our growth rates over small-body size we obtained the minimum mass past which pebble accretion timescales drop below the lifetime of the gas disk, $M_{\rm{peb}}$. By assuming that the early stages of growth are set by gravitational focusing of planetesimals, we were able to translate these minimum masses into limits on the semi-major axes where gas giant growth is possible. We demonstrated that as the disk becomes more turbulent, the range of semi-major axes where gas giants can grow is sharply reduced. These effects may play a large role in the paucity of gas giants at wide orbital separations found by direct-imaging surveys; if disks are not quiescent enough, then pebble accretion may simply produce smaller planets that are unable to accrete sufficient mass in small bodies to go critical. In addition, our mass limits are relevant regardless of how early growth proceeds -- for example, if a body were scattered from the inner disk and it exceeded our minimum mass, it could grow to $M_{\rm{crit}}$ and trigger runaway gas accretion. We also presented approximate analytic expressions for both $M_{\rm{peb}}$ and the upper distance limit where gas giants can form, $a_{\rm{upper}}$.

In addition to the strength of turbulence, we find that the available particle sizes and abundance of planetesimals are major factors in where gas giants can form. Gas-assisted growth is sensitive to the Stokes numbers of the pebbles, as opposed to their absolute size. Thus, if particles of the ``correct" range of Stokes numbers are not available, then gas-assisted growth timescales can be quite slow. Furthermore, if planetesimals are not abundant enough, then the early stages of growth via planetesimal accretion can take too long for subsequent growth via pebble accretion to occur on rapid timescales. 

Finally, we examined the role that viscosity plays in determining the final mass that gas giants reach, in addition to setting where a critical mass core can form. We find that, regardless of the quantitative metric used to determine the final gas giant mass, higher-viscosity disks should feature higher-mass gas giants but at smaller orbital separations. More quantitatively, at the lower $\alpha$ values needed to produce gas giants out to $a \gtrsim 70$ AU, the gas giants formed will be too low-mass to have been observed by direct-imaging surveys. Thus, there may lurk a population of wide orbital separation gas giants that the current generation of imaging surveys has yet to detect.

Thus, if growth of gas giants at wide orbital separations proceeds by gas-assisted growth, and if gap opening sets the final masses of gas giants, we can make qualitative predictions about the observed population of gas giant planets. For a given stellar mass, we would expect that higher-mass gas giants should be observed closer in to the central star, as the disks with higher levels of turbulence will produce higher-mass gas giants at smaller orbital separations. We note that this conclusion may be altered if final planet masses depend on disk surface density, which is not the case for the gap-opening criterion we use. For different stellar masses, we expect higher-mass stars to exhibit higher levels of ionizing radiation. Thus, larger stars may have disks with higher $\alpha$ values and consequently host more massive gas giants.  In addition, the larger disk masses exhibited by more massive stars could push the limits for growth past the distances given here for our fiducial surface density, allowing planets to form at larger distances in high-$\alpha$ disks.  We suggest that this propensity to produce massive planets in high-turbulence disks may be the reason that most currently observed directly imaged gas giants have been found orbiting A stars. 

\vspace{1mm}

\noindent The authors wish to thank Diana Powell, Renata Frelikh, and John McCann for useful discussions, and Eugene Chiang for his thoughtful suggestions on the manuscript. We also thank the anonymous referee for their helpful comments that improved the quality of the manuscript. MMR and RMC acknowledge support from NSF CAREER grant number AST-1555385.

\appendix

In this Appendix, we describe in more detail how we calculate the parameters that are used to determine $t_{\rm{grow}}$. For more information, see R18.

\section{}

\subsubsection*{Drag Forces}

We take the drag force to be divided into two regimes, depending on the size of the small bodies $r_s$: for $r_s < 9 \lambda / 4$, the particle is in the diffuse regime, and for $r_s > 9 \lambda / 4$, the particle is in the fluid regime. Here $\lambda = \mu / (\rho_g \sigma)$ is the mean free path of gas particles and $\sigma$ is the neutral collision cross section. In the diffuse regime, the drag force is given by the Epstein drag law,
\begin{align} \label{eq:epstein}
F_{D, \, \text{Epstein}} = \frac{4}{3}\pi\rho_gv_{th} v_{\rm{rel}} r_s^2 \; ,
\end{align}
where $\rho_g$ is the density of the gas, $v_{th} = \sqrt{8/\pi} c_s$ is the gas's average thermal velocity, and $v_{\rm{rel}}$ is the relative velocity between the small body and the gas.

In the fluid regime, the drag force is given by (\citealt{cheng_drag})
\begin{align} \label{eq:cheng_drag}
F_{D,\text{smooth}} = \frac{1}{2} C_D  (Re) \pi r_s^2 \rho_g v_{\rm{rel}}^2 \; ,
\end{align}
where $Re \equiv 2 r_s v_{\rm{rel}} / \left(0.5 \, v_{th} \lambda \right)$ is the Reynolds number of the particle and
\begin{align}
C_D(Re) = \frac{24}{Re} \left ( 1 + 0.27 Re \right)^{0.43} + 0.47 \left [ 1 - \exp \left( - 0.04 Re^{0.38} \right) \right] \; ,
\end{align}
which is used to smooth between the Stokes and RAM regimes. Note that for small $Re$, the smoothed drag force reduces to the Stokes drag law,
\begin{align}
F_{D, \, \text{Stokes}} = 3 \pi \rho_g v_{th} v_{\rm{rel}} \lambda r_s \; ,
\end{align}
which depends linearly on $v_{\rm{rel}}$.

Using $F_D$, we can calculate the stopping time $t_s \equiv m v_{\rm{rel}} / F_D$. In the Epstein regime, the drag force is independent of $v_{\rm{rel}}$, and in the fluid regime, it is solved for numerically using Equations \eqref{lam_rel_gas} and \eqref{turb_rel_gas} below for $v_{\rm{rel}}$.

\subsubsection*{Velocity of Small Bodies}

We will use $k$ and $g$ to denote velocities relative to the local Keplerian velocity and relative to the gas, respectively, as well as $\ell$ and $t$ to denote the velocities due to the interactions with the laminar and turbulent components of the gas velocity.

For a purely laminar gas flow, \cite{nag} gave the relative velocity between a small body of stopping time $\tau_s \equiv t_s \Omega$ and the sub-Keplerian gas as

\begin{align} 
v_{pg,\ell}=-2\eta v_{k} \left[ \frac{\tau_s}{1 + \tau_s^2} \right] \hat{\boldsymbol{r}} - \eta v_{k} \left[ \frac{1}{1+\tau_s^2} - 1 \right ] \hat{\boldsymbol{\phi}} \label{lam_rel_gas}
\end{align} 
where $\eta \equiv c_s^2/\left(2 v_k \right)$ is a measure of the pressure support in the disk, and $v_k \equiv a \Omega$ is the local Keplerian velocity. In the presence of turbulent gas velocity, we need to modify the above equations by introducing additional terms to account for the particle's interaction with turbulence.

In order to parameterize the strength of the turbulence, the turbulent gas is described in terms of the Shakura and Sunyaev $\alpha$ parameter (\citealt{ss_alpha}), which is defined by $\nu_t = \alpha c_s H_g$, where $\nu_t$ is the effective kinematic viscosity of the turbulent gas defined in terms of the viscous torque in the disk and $H_g = c_s/\Omega$ is the scale height of the gas. In terms of this parameter, we take the RMS turbulent gas velocity to be $v_{\rm{gas},t} = \sqrt{\alpha} c_s$. 

The small bodies will acquire a nonzero RMS velocity due to their interactions with the turbulence. The strength of this interaction is generally parameterized by the particles' Stokes number, $St \equiv t_s/t_L$, where $t_L$ is the overturn time of the largest-scale eddies. For our purposes, we take $t_L \sim \Omega^{-1}$, in which case $St = \tau_s$, and the particle's interaction with the gas can be expressed in terms of only one dimensionless parameter. Hereafter, we will exclusively refer to this parameter as $St$. For the drift velocity (relative to the local nebular gas velocity) of the particles due their interactions with the turbulent gas velocity, we use

\begin{align} \label{turb_rel_gas}
v_{pg,t}=\begin{dcases*}
\sqrt{\alpha} c_s\left(\frac{St^2 \left(1-Re_t^{-1/2} \right)}{(St+1) \left(St+Re_t^{-1/2} \right)}\right)^{1/2} & $St < 10$\\
\sqrt{\alpha} c_s \sqrt{\frac{St}{1+St}} & $St \geq 10$
\end{dcases*}
\end{align}
where $Re_t \equiv \nu_t/(v_{th}\lambda)$ is the Reynolds number of the turbulence. The former expression comes from \cite{ch03}, who derived approximate analytic expressions for the RMS velocity. While they derived their results for $St \ll 1$,  \cite{oc07} argued that these expressions hold to order unity for particles of arbitrary size. The second expression is based on order-of-magnitude arguments (see, e.g. \citealt{cy10}) and is included to give the expected behavior that $v_{pg,t} \rightarrow v_{\rm{gas},t}$ as $St \rightarrow \infty$.

The presence of a laminar velocity has effects on the turbulent velocity of the particle as well. To incorporate this effect, we use an effective eddy turnover time given by (\citealt{orb})

\begin{align}
t_L=\frac{\Omega^{-1}}{\sqrt{1+(v_{pg,\ell}/v_t)^2}} \label{eddy_cross_eqn} \; .
\end{align}

The velocity of particles relative to the gas is necessary for drag calculations, but the velocity of the particles relative to Keplerian is also needed for calculating what sets the particles' approach velocity, $v_\infty$. The above velocities relative to the local Keplerian velocity are

\begin{align}
v_{pk,\ell} = -2\eta v_{k} \left[ \frac{\tau_s}{1 + \tau_s^2} \right] \boldsymbol{\hat{r}}
-\eta v_{k} \left[ \frac{1}{1+\tau_s^2}\right ] \boldsymbol{\hat{\phi}}
\end{align}
for the laminar component and 

\begin{align} \label{eq:v_turb_kep}
v_{pk,t}=\begin{dcases*}
\sqrt{\alpha} c_s\left(1-\frac{St^2 \left(1-Re_t^{-1/2} \right)}{(St+1) \left(St+Re_t^{-1/2} \right)}\right)^{1/2} & $St < 10$\\
\sqrt{\alpha} c_s \sqrt{\frac{1}{1+St}} & $St \geq 10$
\end{dcases*}
\end{align}
for the turbulent component. The total RMS velocity in both frames is

\begin{align}
v_{pg} &= \sqrt{v_{pg,\ell}^2 + v_{pg,t}^2}\\
v_{pk} &= \sqrt{v_{pk,\ell}^2 + v_{pk,t}^2},
\end{align}

The other relevant velocity in our calculation is the shear velocity, which arises due to the differential rotation in the Keplerian disk. For two bodies separated by a radial distance $r$, the relative velocity is approximately

\begin{align}
v_{\rm{shear}} \approx r \Omega \; .
\end{align}

We take the relevant velocity for calculating $v_\infty$ as the maximum of the drift velocity due to gas interactions and the shear velocity

\begin{align}
v_\infty = \max(v_{pk},v_{\rm{shear}}) \; .
\end{align}

During the encounter, the particle's velocity may be changed substantially due to its interaction with the core. If $v_{\rm{orbit}} = \sqrt{G M / R_{\rm{acc}}}$, then for $v_\infty < v_{\rm{orbit}}$ we expect the particle to be accelerated up to $v_{\rm{orbit}}$, while for $v_\infty > v_{\rm{orbit}}$ we expect the particle's velocity to be perturbed by an amount $v_{\rm{kick}} = G M / (R_{\rm{acc}} v_\infty)$ perpendicular to the direction of approach.  Therefore, the velocity of the small body during its encounter with the core is given by
\begin{align}
v_{\rm{enc}} &= \max\left(v_{\rm{orbit}}, v_{pg}\right); v_\infty < v_{\rm{orbit}}\\
v_{\rm{enc}} &= \max\left(v_{\rm{kick}}, v_{pg}\right); v_\infty > v_{\rm{orbit}}
\end{align} 

\subsubsection*{Length Scales} \label{lens}

In order to determine $R_{\rm{acc}}$ we first need to calculate $R_{\rm{stab}}$, the radius at which the small body can stably orbit the core. While $R_H$ can set this length scale, for small bodies we also need to determine if gas drag will shear bodies off the core. To include this effect, we use the WISH radius defined by \cite{pmc11}, which is the radius at which the differential acceleration from gas drag between the two bodies is balanced by the mutual gravitational acceleration, that is,

\begin{align} \label{eq:app_r_ws_full}
R_{WS}^\prime=\sqrt{\frac{G(M+m)}{\Delta a_{WS}}} \; ,
\end{align}
where 

\begin{align} \label{DiffAcc}
\Delta a _{WS} = \left| \frac{F_D(M,v_{\rm{rel}})}{M} - \frac{F_D(m,v_{\rm{rel}})}{m} \right| \; ,
\end{align}
where $v_{\rm{rel}}$ is the velocity of the gas relative to the particle. Note that for our purposes, $M \gg m$, so the first term in \eqref{DiffAcc} is generally negligible. 

To calculate $v_{\rm{rel}}$, we note that during the encounter, the small body will be decelerated by the core, bringing the small body's velocity close to the core's velocity. In the most extreme case, the small body will temporarily experience the full gas velocity relative to the core's velocity. Thus, the correct velocity for calculating the drag force in our expression for $R_{WS}^\prime$ is the velocity of the gas relative to the orbital velocity of the core. We approximate this velocity as $v_{\rm{rel}} \approx \max(v_{\rm{gas}},v_{\rm{shear}})$
where $v_{\rm{gas}} = \sqrt{\eta^2 v_k^2 + \alpha c_s^2}$ is the RMS total gas velocity relative to the local Keplerian velocity, and $v_{\rm{shear}} = R_{\rm{acc}} \Omega$ is the shear velocity between the gas at impact parameter $R_{\rm{acc}}$ and the core. In the case where $v_{\rm{gas}} > v_{\rm{shear}}$ we refer to the impact parameter as $R_{WS}$ (without the prime), which has the approximate value,
\begin{align}
R_{WS} \approx \sqrt{\frac{G M t_s}{v_{\rm{gas}}}} \; .
\end{align}
For the case where $v_{\rm{shear}} > v_{\rm{gas}}$, Equation \eqref{eq:app_r_ws_full} must be solved for the impact parameter, as the right-hand side of the equation now depends on $R_{WS}^\prime$. In this regime we refer to the impact parameter as $R_{\rm{shear}}$; for a particle in a linear drag regime with $M \gg m$, the value of $R_{\rm{shear}}$ is well approximated by
\begin{align}
R_{\rm{shear}} \approx R_H \left(3 St \right)^{1/3} \; .
\end{align}
Then, $R_{\rm{stab}}$ is computed as 

\begin{align}
R_{\rm{stab}} = \min(R_{WS},R_{\rm{shear}},R_H),
\end{align}
and the relevant length scale for the timescale calculation, $R_{\rm{acc}}$, is

\begin{align}
R_{\rm{acc}} = \max(R_{\rm{stab}},R_{\rm{atm}}).
\end{align}
where $R_{\rm{atm}} = \min(R_b,R_H)$. For the range of core masses we consider in this work, we have $R_b <  R_H$. Since pebble accretion shuts off for $R_{WS} < R_b$ (see Section \ref{part_size}), we always have $R_{\rm{acc}} = R_{\rm{stab}}$ in this regime.

The particle scale height can be set by either the Kelvin-Helmholtz shear instability (\citealt{kh}, R18)

\begin{align}
H_{KH} = \frac{H_g^2}{a} \min(1,St^{-1/2}) = \frac{2\eta v_k}{\Omega} \min(1,St^{-1/2}) \label{HKH} \; ,
\end{align}
or by turbulence (see, e.g. \citealt{dub}, \citealt{carb}),

\begin{align}
H_t = \min \left(\sqrt{\frac{\alpha}{St}} H_g,H_g \right) \; .
\end{align}
The particle scale height is then

\begin{align}
H_p = \max(H_{KH},H_t) \; .
\end{align}

The relevant size of the height of the accretion ``rectangle," $H_{\rm{acc}}$, depends on the relative size of $R_{\rm{acc}}$ and $H_p$. If $H_p < R_{\rm{acc}}$ then accretion is cut off at $H_p$; if $H_p > R_{\rm{acc}}$ then the body can accrete over the entirety of $R_{\rm{acc}}$ in the vertical direction as well as the horizontal. We therefore take

\begin{align}
H_{\rm{acc}} = \min(H_p,R_{\rm{acc}}) \; .
\end{align}

\subsubsection*{Growth Timescale}
The above parameters are used to calculate the growth timescale using the formula
\begin{align} \label{eq:app_t_grow}
t_{\rm{grow}} = \frac{M H_p}{2 f_s \Sigma v_\infty R_{\rm{acc}} H_{\rm{acc}}} \; ,
\end{align}
where $\Sigma$ is the surface density of the nebular gas and $f_s$ is the solid-to-gas mass ratio. Particles that shear into $R_H$ undergo chaotic trajectories, which can lead to extended interactions with the core. To capture this effect, we treat accretion probabilistically in this regime, with particles having accretion probability $P = \min \left(1,W/KE \right)$. Thus, for particles that have $R_{\rm{stab}} = R_H$ and $v_\infty = v_H$, we have
\begin{align}
t_{\rm{grow}} = \frac{t_{\rm{grow}}^\prime}{\min(1,W/KE)} \; ,
\end{align}
where $t_{\rm{grow}}^\prime$ is given by Equation \eqref{eq:app_t_grow}. See R18 for more details.

\section{}

\bibliographystyle{yahapj}
\bibliography{gas_bib}

\end{document}